# Progress in nonmagnetic impurity doping studies on Fe-based superconductors

Jun Li,[1,2,3*] Yan-Feng Guo,[4,2,†] Zhao-Rong Yang,[5,6] Kazunari Yamaura,[2,7]

Eiji Takayama-Muromachi,[2] Hua-Bing Wang,[1,2,3] Pei-Heng Wu [1]

[1] Research Institute of Superconductor Electronics, Nanjing University, Nanjing 210093, China

[2] National Institute for Materials Science, Tsukuba 305-0044, Japan

[3] Cooperative Innovation Centre of Terahertz Science, Chengdu 610054, China

[4] School of Physical Science and Technology, ShanghaiTech University, Shanghai 200031, China

[5] Key Laboratory of Materials Physics, Institute of Solid State Physics, Chinese Academy of Sciences, Hefei, Anhui, 230031, China.

[6] Collaborative Innovation Center of Advanced Microstructures, Nanjing University, Nanjing, 210093, China

[7] Graduate School of Chemical Science and Engineering, Hokkaido University, Hokkaido 060-0810, Japan

Corresponding authors: J.L. (E-mail: junli@nju.edu.cn), Y.F.G.
(E-mail:guoyf@shanghaitech.edu.cn)

## Abstract

We review the progress of nonmagnetic impurity doping study on the Fe-based superconductors. On the theoretical side, two highly potential candidates for the pairing symmetry order parameter, i.e. the multi-gap $s_{++}$ and $s_{\pm}$ wave models, have been proposed but continuously debated. The debate arises because of the complex gap structure, exceptive magnetic and metallic behaviors of Fe-based superconductors which may make the influences of nonmagnetic defects varied in chemical potential, impurity disorder, inter- and intra-band scattering strength, and electron localization, and hence difficulty in directly obtaining the most important information for understanding the symmetry order parameter. Experimentally, the nonmagnetic impurity substitution study has been widely carried out, which has provided very useful insights. We review herein the various nonmagnetic impurity doping experiments, including the controlled defects within the superconducting $Fe_2X_2$ planes through samples quality improvement, a single impurity effects on the electronic state and local moment, the magnetic response of the $Fe_2X_2$ planes both on macroscopic scale as antiferromagnetic state and local scale of moment, as well as the significant effect in modifying the transport properties. The experiments enable us to qualitatively analyze the nonmagnetic impurities effects on the superconducting state for many Fe-based superconductors. We also propose herein some strategies for the nonmagnetic impurity doping study. As an important model for explaining the nonmagnetic impurities doping effects, the pair-breaking model is compared with various theoretical approaches via analyzing the pair-breaking rates of various Fe-superconductors.

# 1. Introduction

In 2006, Kamihara *et al*. discovered superconductivity with a critical temperature ($T_c$) of 5 K in LaFePO$_{1-x}$F$_x$ which has a layered ZrCuSiAs-type (1111-type) tetragonal structure [1]. Adopting the same strategy by substituting fluorine for oxygen in an analogue LaFeAsO, a much higher $T_c$ of 26 K was achieved in early 2008 [2]. The exciting discovery aroused another research climax of superconductivity which announced that a new high-$T_c$ superconductor family, the so-called Fe-based superconductor (FBS), was born. The FBS family has attracted intensive attentions and the studies have been advanced rather rapidly [1-7]. This is not only due to that the FBS is a second class of high-$T_c$ superconductors after the cuprate superconductors, but also because it is highly promising to understand the superconducting (SC) mechanism of high-$T_c$ by comparing the two families.

To date, one of the remained key issues is to elucidate the pair-symmetry of the FBS, for which several possible models were proposed just after the discovery [8-18]. Among the various models, the multi-gaped *s*-wave is generally acceptable but still has been debated between the $s_\pm$ [8, 9] and $s_{++}$ wave models [10, 11]. In both models, the Fermi surface has the same hole-type pockets but the electron-type pockets within the $s_\pm$ wave model have opposite signs, i.e. a sign-reversal *s*-wave model, whilst have the same sign within the $s_{++}$ wave model. Additionally, the *d*-wave model with opposite signs for the nearest-neighbor electron pockets remains competing, provided that there are nodes on the hole pockets or even on both the electron and hole pockets [10, 16, 17]. The various experiments results can hardly reach a consensus on the symmetry model. More recent results even suggested that different systems in the iron-pnicitide family may represent different pairing symmetries, and the superconductors with different doping levels are as well [12,15,18]. The divergences appeal further investigations to provide solid evidences. Among the various strategies, the impurity substitution, especially the nonmagnetic impurities substitution for Fe, is one of the most promising approaches to address the issue.

The SC Fe$_2$X$_2$ (*X*=As, P or Se) plane, which is similar as the SC CuO$_2$ layer in cuprate superconductors, is a common feature shared by the various iron-pnicitide superconductors. The substitution of point defects for the Fe-site has been proposed for understanding the physical properties, which has been demonstrated as an effective way and has been intensively undertaken in cuprate superconductors. According to Anderson's theorem [19], the nonmagnetic impurity (NMI) cannot break Cooper pairs in an isotropic SC gap but for an anisotropic gap, while the pair-breaking effect of the magnetic impurities is independent of gap type. Thus, the introduction of nonmagnetic point defects is crucial to probe the information of the gap. The zinc ion ($Zn^{2+}$) with tightly closed *d*-shell is considered as an ideal NMI [20-28]. The study of Zn substitution effects for Cu on the cuprate family such as $YBa_2Cu_3O_{7-\delta}$ [20-22], $(La,Sr)_2CuO_4$ [20, 23-25], and $Bi_2Sr_2CaCu_2O_8$ [20, 26-28] was carried out over the last two decades. Only a few at.% of Zn which acts as a strong scattering center can remarkably depress SC due to the *d*-wave anisotropic gap [20]. Since the doped Zn often plays a crucial role in pairing symmetry judgment demonstrated by the investigations on many superconductors, it is expected to work in the FBS as well.

In this article we review both theoretical and experimental studies of the NMI effects on the superconductivity of FBS. We start with the introduction of the potential gap symmetries of FBS and recent theoretical studies on the defect impurities in **Section 2**, and then overview the NMI substitution methods in **Section 3**, including both chemical doping and irradiation inducing defects. The effects of NMI on the various properties of the host superconductors are reviewed by introducing the various experiments in **Section 4**: (i) The sample quality will be discussed for the controlled impurities within the SC $Fe_2X_2$ planes in **Section 4.1.** (ii) In **Section 4.2**, a single impurity effects on the electronic state and local moment are introduced. (iii) Magnetic response of the $Fe_2X_2$ planes to NMI on both macroscopic scale as antiferromagnetic state and local scale of moment will be reviewed in **Section 4.3**. (iv) Significant modifications of the transport properties of the superconductors are discussed in details in **Section 4.4**. (iv) Some potential experiments for introduction of NMI are also proposed in **Section 4.5**. (v) Finally, as the most prominent qualitative feature from the impurities doping, the effects on SC state are discussed in **Section 4.5**, which covers many systems of superconductors. As a summary, the pair-breaking rates in different systems will be reviewed in **Section 5** by comparing with various theoretical approaches.

## 2. Theoretical background

### 2.1. Gap symmetry of Fe-based superconductors

Theorists have proposed four models as candidates for the SC energy gap symmetry of FBS as shown in **Figure 1** [6]. **Figure 1 (a)** illustrates the order parameter (OP) for the conventional Bardeen-Cooper-Schrieffer (BCS) superconductors [29], in which the electrons pairing via gaining energy from the electron-phonon interactions and the SC OP is the same for all electrons, namely, an "*s*-wave" model. The *s*-wave SC is associated with an isotropic energy gap for the Cooper pairs, being completely symmetric along all directions. The electron-electron magnetic interaction [30-34] dominates another type of SC gap with the "$d$-wave" OP as shown in **Figure 1(b)**, where the sign variation of the OP has direction dependence, e.g. $\cos 2\alpha$, which can result in an anisotropic gap symmetry. The OP of the cuprate superconductors is generally of the $d$-wave symmetry. Additionally, the multi-gap model ($s_{++}$ wave) is also proposed, as shown in **Figure 1(c)** [35]. It is believed that the SC mechanism for $MgB_2$ is of the BCS-type with the $s_{++}$ symmetry, whereby the Fermi surface comprises two small hole pockets around the **Γ** = (0, 0) point and two electron pockets around the **M** = (π, π) point in the 2-Fe (Mg?) Brillouin zone. The OP is positive for all bands but has different magnitudes for different bands, belong to an isotropic symmetry. After the discovery of FBS, another type of multi-gap model, the so-called "$s_{\pm}$ wave" model (see **Figure 1 (d)**) was proposed to interpret the OP. In the $s_{++}$ wave model, the Fermi surface also consists of two small hole pockets and two electron pockets, whereas the OP in some band(s) has positive sign but in the other band(s) has negative sign. Within such a framework, the $s_{\pm}$ wave gap structure possesses a strongly anisotropic symmetry.

For FBS, the conventionally simple $s$-wave model has already been eliminated by various studies and the multi-gapped $s$-wave models, $i.e.$, the $s_{\pm}$ [8,9] and the $s_{++}$ wave [10-12], are currently promising candidates, where the former likely has a magnetic fluctuation origin while the latter likely has a charge fluctuation origin. Both models have positive OPs for the hole Fermi pockets, while have opposite signs for the electron pockets, thus the $s_{\pm}$ wave is sometimes called as a sign-reversal $s$-wave, and the $s_{++}$ state as a non-sign-reversal. As a competing candidate, the $d$-wave model with nodes on the hole pockets or even on both the electron and hole pockets remains possible for the OP of FBS [10,16,17].

Various experiments have been performed to investigate the OP symmetry of FBS, including angle-resolved photoemission spectroscopy, quantum oscillation, London penetration depth, and Josephson tunneling effect, and so on. However, the issue remains inclusive because of the rather contrary results derived from these experiments. Here, we will review the NMI doping study on this issue.

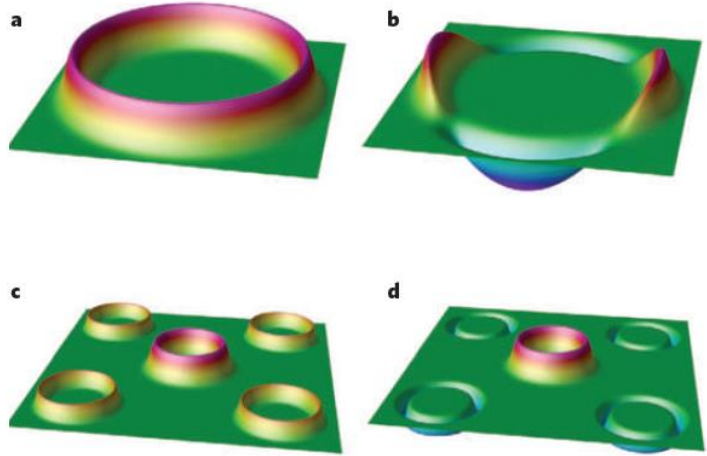

**Figure 1.** Schematic images for various SC order parameter models. (a) single-gap *s*-wave for the conventional superconductors, (b) single-gap *d*-wave for the cuprates, (c) multi-gap $s_{++}$ wave for MgB$_2$, and (d) multi-gap $s_{\pm}$ wave, from Mazin [8].

## 2.2. Nonmagnetic impurity doping effect on superconductivity

In this section, we focus on theoretical studies on the effect of impurity atoms or the atomic defects. The pair-breaking effect by magnetic and NMI will be introduced. As we mentioned above, magnetic impurity suppresses SC independently of the gap symmetry while the NMI effect depends strongly on it. The NMI, for instance Zn, will suppress $T_c$ according to the Abrikosov-Gor'kov (AG) formula [36-38]

$$\ln\left(\frac{T_{c0}}{T_c}\right) = \Omega[\varphi(\frac{1}{2} + \frac{\mu}{2}) - \varphi(\frac{1}{2})] \qquad (1),$$

where $\varphi(x)$ is the digamma function. In mathematics, the digamma function is defined as the logarithmic derivative of the gamma function [37]

$$\varphi(x) = \frac{d}{dx} \ln \Gamma(x) = \frac{\Gamma'(x)}{\Gamma(x)} \qquad (2).$$

It also has an integral representation as

$$\varphi(x) = \int_0^\infty (\frac{e^{-t}}{t} - \frac{e^{-xt}}{1 - e^{-t}})dt \qquad (3).$$

The $\mu$ in Eq. (1) is defined by

$$\mu = \frac{\hbar}{2\pi k_B T_c \tau} \qquad (4),$$

where $\tau$ is the relaxation time of NMI scattering. $\Omega$ in Eq.(1) is the gap anisotropy defined as

$$\Omega \equiv 1 - \frac{<\Delta(k)>^2}{<\Delta(k)^2>} \qquad (5).$$

For an isotropic SC gap, for instance the isotropic *s*-wave, $\Omega=0$, while $\Omega$ is 1 for an anisotropic gap like *d*-wave and anisotropic *s*-wave.

As we can see from **Eq**. (1), the introduction of NMI does not break paired electrons for an isotropic SC gap but for an anisotropic gap [38]. However, $T_c$ also depends on the Debye frequency $\omega_D$ and density of states $D(E_f)$ according to the BCS theory in a form

$$T_c \approx \frac{1.14\hbar\omega_D}{k_B} exp(\frac{-1}{D(E_f)V}) \qquad (6),$$

where $V$ is the electron-phone interaction strength. If the introduction of NMI affects $D(E_f)$ (due to the carrier doping) or $\omega_D$, $T_c$ will consequently vary even in the isotropic *s*-wave superconductors. Except for (?) the effect on $D(E_f)$ and $\omega_D$, NMI can also break the Cooper pairs in anisotropic superconductors due to $\Omega\neq0$, as given by the following equation derived from **Eq**. (1)

$$\frac{T_c}{T_{c0}} = 1 - \frac{\pi\hbar}{8k_B}\frac{1}{\tau} \qquad (7).$$

It is clear from Eq.(7) that the suppression of $T_c$ is proportional to the impurity scattering strength and the last part is defined as pairing breaking rate

$$\alpha = \frac{\pi\hbar}{8k_B}\frac{1}{\tau} \qquad (8),$$

where $1/\tau_c$, which is defined as the critical scattering rate where SC is completely suppressed, can be expressed as

$$\frac{1}{\tau_c} = \frac{8k_B}{\pi\hbar} \qquad (9).$$

We will strengthen the discussions on this critical parameter in **Section 3 and 4**.

## 2.3. Pair-breaking

To study the pair-breaking effect from magnetic and NMI on FBS, one of the most direct ways is to quantify the relation between $T_c$ suppression and the doping level $n_{imp}$. Onari and Kontani [11] firstly analyzed the effect of local impurity on iron pnicitides, based on the five-orbital model. **Figure 2** shows the calculation results for $T_c$ suppression as a function of doping level for both $s_\pm$ and $s_{++}$ wave models. They suggested that in the $s_\pm$ wave state, the inter-band impurity scattering is promoted by the $d$-orbital degree of freedom. Consequently, the $s_\pm$ wave state should be very fragile against impurities, regardless of magnetic or nonmagnetic impurities. $T_c$ was estimated to vanish with $n_{imp}$ of 0.01, 0.02, and 0.066 for $I = 1$ eV, $\infty$, and -1 eV, respectively, thus giving rise to the reduction rate of $T_c$ ($dT_c/dn_{imp}$) caused by the impurity is ~$50z$ K/at.% [11], where $z$ is the renormalization factor (= $m/m^*$; $m$ and $m^*$ are the band-mass and the effective-mass, respectively). Since the effective mass was estimated to be between $2m_e$ and $4m_e$ in the 122-type superconductors from ARPES measurements [39-41], one can obtain $dT_c/dn_{imp}$= 25 K/at.% (17 K/at.%) for $z = 0.5$ ($z = 0.33$). As a contrast, non-sign reversal $s_{++}$ wave state can strongly against the NMI. It seems not so complicated to judge which order parameter symmetry is the nature of these superconductors, once we can quantify the relation of $T_c$ vs. $n_{imp}$. Nevertheless, direct accurate determinations of the real scattering rate are hardly achieved, which is generally much more complicated by some other important factors that can also result in obvious $T_c$ suppression.

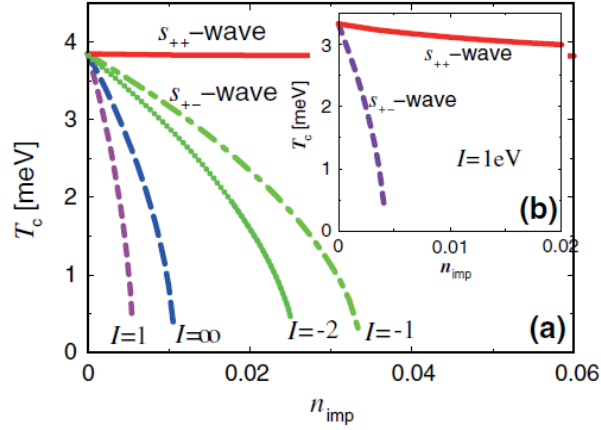

**Figure 2**. Obtained $T_c$ as functions of $n_{imp}$ for the case of $s_{++}$ wave and $s_{\pm}$ wave states with various chemical potential. Here the $T_{c0}$ with impurity-free was considered as 46 K, from the reference by Onari and Kontani [11].

The impurity disorder can also affect the SC due to that the SC mostly origins from doping. Efremov *et al*. [14] analyzed the effect from disorder on two-band gap symmetry, namely, intra- and inter-band impurity scattering for which the $T_c$ suppression rate $\widetilde{\Gamma}_{ab}$ is given by

$$\widetilde{\Gamma}_{ab} = \Gamma_{a(b)} \frac{(1-\widetilde{\sigma})}{\widetilde{\sigma}(1-\widetilde{\sigma})\eta \ (N_a + N_b)^2 / N_a N_b + (\widetilde{\sigma}\eta - 1)^2} \quad (10).$$

Here, $\eta$ is the parameter as the ratio of intra- and inter-band scattering as $\eta = v^2 / u^2$, $N_a$ and $N_b$ are the density of states for each band ($a$, $b$) at the Fermi level, and $\widetilde{\sigma}$ and $\Gamma_{a(b)}$ are the generalized cross-section normal state scattering rate parameters given as follow:

$$\widetilde{\sigma} = (\pi^2 N_a N_b u^2) / (1 + \pi^2 N_a N_b u^2) \quad (11),$$

and

$$\Gamma_{a(b)} = n_{imp} \pi N_{a(b)} u^2 (1 - \widetilde{\sigma}) \quad (12).$$

In the weak scattering (Born) limit, one can get $\widetilde{\sigma} \to 0$, while in the unitary limit (strong scattering) $\widetilde{\sigma} \to 1$. On the other hand, for the strong scattering case, $\widetilde{\Gamma}_{ab} \to 0$, hence $T_c$ is independent of NMI.

**Figure 3** gives the calculation results for the effective inter-band scattering rate $\widetilde{\Gamma}_{ab}$ dependent $T_c$ suppression in various $\widetilde{\sigma}$ and $\eta$. The slopes of $T_c$ collapse onto one of three 'universal' curves, depending on the average pairing strength parameter $\langle\lambda\rangle$ as positive, zero, and negative. Here the sign of $\langle\lambda\rangle$ depends on the coupling constant ($\lambda_{ij}$) of two-band superconductor as given by

$$\langle\lambda\rangle \equiv (\lambda_{aa}+\lambda_{ab})N_a N^{-1} + (\lambda_{ba}+\lambda_{bb})N_b N^{-1} \tag{13},$$

where $N = N_a + N_b$. The results indicate that a universal behavior of $T_c$ suppression controlled by a single parameter $\langle\lambda\rangle$. As a result, the $s_\pm$ wave SC can be suppressed by NMI for $\langle\lambda\rangle \le 0$, but also demonstrates robust against impurities once $\langle\lambda\rangle > 0$. Typically, the critical value of the scattering rate $\Gamma^{\mathrm{crit}}$ defined by $T_c(\Gamma^{\mathrm{crit}}) = 0$ is given by

$$\Gamma^{\mathrm{crit}}/T_{c0} = \pi/2\gamma \approx 1.12 \tag{14}$$

within the Abrikosov-Gor'kov (AG) theory. In addition, as the increase of such disorder, the $s_\pm$ wave may change to a fully-gaped symmetry, $s_{++}$ wave, which was suggested to be manifested by thermodynamic and transport properties.

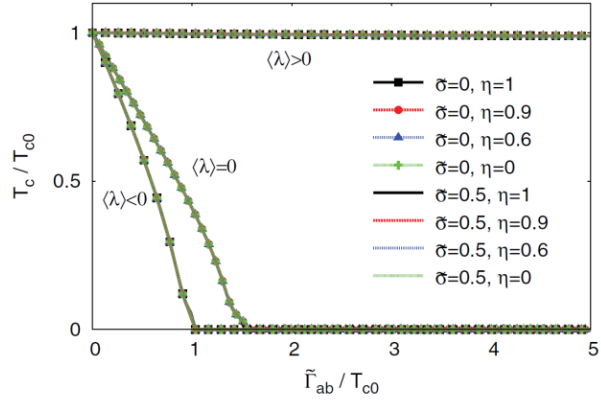

**Figure 3**. Normalized $T_c$ as a function of effective inter-band scattering rate $\widetilde{\Gamma}_{ab}$ for various $\widetilde{\sigma}$ and $\eta$, where the $\widetilde{\Gamma}_{ab}$ was calculated by using the linearized Eliashberg equation. The curves for different sets of $\widetilde{\Gamma}_{ab}$ and $\eta$ overlap and fall onto one of the three universal curves depending on the $\langle \lambda \rangle$. $N_b/N_a$ = 2, coupling constants for illustrative purpose are chosen for $\langle \lambda \rangle > 0$ as ($\lambda_{aa}$, $\lambda_{ab}$, $\lambda_{ba}$, $\lambda_{bb}$) = (3,-0.2,-0.1,0.5), for $\langle \lambda \rangle$ = 0 as (2,-2,-1,1), and for $\langle \lambda \rangle < 0$ as (1,-2,-1,1), from Efremov e*t al*. [14].

On the other hand, Wang *et al*. [42] and Efremov *et al*. [43] also suggested that measurements of $T_c$ suppression relative to the amount of chemical potential and disorder can hardly determine the gap structure in multiband systems. To improve the situation, one first needs to find a way to create points like potential scattering centers, so as to create disordered systems. From an experimental view, we need more accurate measurements on residual resistivity $\rho_0$ from high quality NMI-doped single crystals. Here, $\rho_0$ is corresponding to the change in the extrapolated $T \to 0$ value of the resistivity with disorder. Compared with the $n_{imp}$ dependent $T_c$, the observation of $\rho_0$ is the more accurate way to explore the pair-breaking effect from impurity scattering. Wang *et al*. [42] calculated the relation between $T_c$ suppression and the change of $\rho_0$, and showed how these results change for various types of gap structures and assumptions regarding the impurity scattering. **Figure 4** shows the calculation results for $T_c$ suppression *vs*. the corresponding change in $\rho_0$ for isotropic $s_{\pm}$ wave model and anisotropic one with various values of the inter- to intra-band scattering ratio $\alpha \equiv u/v$. Note that $s_\pm$ gap can be anisotropic with nodes on the electron pockets, whose gap function ($\Delta_e$) can be given as

$$\Delta_e = \Delta_0(1 + r\cos 2\phi) \tag{15},$$

where $\Delta_0$ is angle-independent part of the gap function, $r$ is the anisotropic factor, and $\phi$ corresponds to the angle around the electron pocket. For a fully isotropic $s$-wave gap, $r = 0$, hence the $T_c$ is independent of NMI doping. In respect that a gap which has nodes on the electron pockets, for instance set $r = 1.3$, the results exhibit a wide variety of initial slopes of $T_c$-$\rho_0$, which strongly depends on the scattering character of the impurity, but not just the doping lever itself. Namely, the $T_c$ may be suppressed weakly or sharply by the change of $\rho_0$, for instance, the pair-breaking residual resistivity $\Delta\rho_0^c$ can be up to 1000 $\mu\Omega$ cm once the $\alpha$ is less to 0.2, and even without any suppression for $\alpha = 0$. Previous calculation results also showed $\alpha$-dependent suppression rate [42].

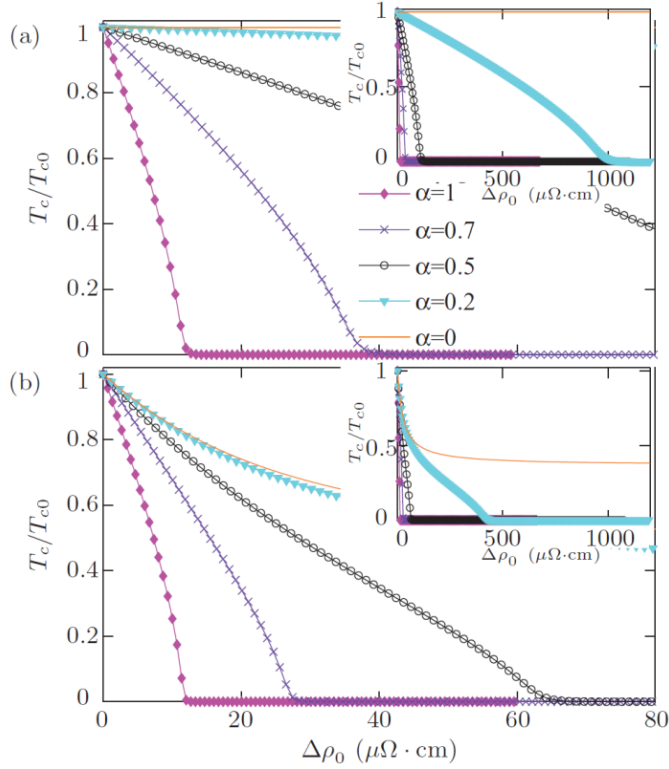

**Figure 4**. (a) Disorder-induced resistivity change $\rho_0$ dependence of normalized critical temperature $T_c/T_{c0}$, for the case of isotropic $s_\pm$ wave pairing with various values of the inter- to intra-band scattering ratio $\alpha \equiv u/v$. Inset: Same quantity plotted over a larger $\rho_0$ scale. (b) The same as (a) but for an anisotropic (nodal) gap with anisotropy parameter $r = 1.3$, from Wang e*t al*. [42].

Studying on the pair-breaking effect associated with the defects is a complicated issue. Besides the pair-breaking and the scattering character of the impurity, one has to consider the separation of the exploration of impurity effects on different length scales, from lattice spacing to the coherence length to sample size, as suggested by Anderson theorem [19]. In addition, Sato *et al*. [44] proposed two other mechanisms for $T_c$ suppression from impurities substitution: (i) the electron localization with sheet resistance $R_\square$ exceeding $h/4e^2 = 6.45$ k$\Omega$, and (ii) the disappearance (or reduction in the area) of the whole Fermi surfaces around the $\Gamma$ point in the reciprocal space. Furthermore, considering the NMI in the isotropic $s_{++}$ wave state, the SC suppression may be caused by the following mechanisms: (i) suppression of the orbital fluctuations, which is a possible origin of the $s_{++}$ wave state, because of the violation of the orbital degeneracy near the impurity atom, and (ii) the strong localization effect in which the mean-free-path is comparable to the lattice spacing.

To well probe pair-breaking behavior, the fabrication of high quality nonmagnetic Zn-doped compound, especially for single crystals, is the most important issue. In the next section, we will show the recent progresses on the fabrication of impurity-substituted FBS.

# 3. Impurity substitution methods

Substitution of impurity is considered as the simplest defect to "tune" the nature (magnetic and/or electronic properties) of the perturbation potential. Generally, the substitution was obtained by either chemical doping during the synthesis process or irradiating defects into a host crystal. Here, the substituted chemicals work as external impurities, while the irradiating defects as intrinsic impurities. In this section, we will discuss the fabrication of chemicals doped FBS crystals using ambient and high-pressure methods, and also impurities introduced by irradiation technique.

## 3.1. Chemical substitution

Compared with the host atoms, the impurities may vary in the atomic size, bond valance, electron structure, magnetic state, and so on. The experimental design for chemical substitution should choose at least the similar atomic size and bond valance of the impurity as the host atoms, and the atomic substitution site should be confirmed as well before and after the synthesis. Taking the well-studied cuprate superconductors as an example [20-28], impurities of Zn, Ni and Co can be successfully substituted for Cu-site owing to the similar atomic size. However, the substitution sites may occur on various Cu-site, for instance in the multi-layered $Bi_2Sr_2CaCu_2O_{8+\delta}$ superconductors the impurities may go into the $CuO_2$-planes or the chain Cu-site, where the former is believed as the SC layer but the latter is non-SC site and will provide misleading on the impurity-doping experiments. Therefore, we have to consider the structure of the FBS before doping.

Up to now, seven series have been discovered in the Fe-based family as given in **Figure 5** [1-8]. Similar to the cuprates, each Fe-based compound has the distinct layered structure with the key layer of Fe$X$ ($X$ =As, P, Te, or Se) (see **Figure 6**). Actually, the common Fe$X$ layer should be more accurately written as Fe$_2X_2$, because half of the $X$ atoms lie above the Fe plane and the others lie below the plane [45]. For the Fe-site, however, we can consider it as only one site regardless of the spin direction. Thus, it consists of a square lattice of iron atoms coordinated tetrahedrally by the $X$ anions that form a checkerboard pattern below and above the Fe plane which can double the unit cell size. Except for the case of $X$=Se, the Fe$_2X_2$ layers are separated by blocking layer(s) through which carrier doping into the Fe$_2X_2$ layer takes place in the same manner as in the cuprates. The blocking layer provides a quasi-two-dimensional character to the crystal since it forms bonds of more ionic nature with the Fe$_2X_2$ layer. The Fe$_2X_2$ layer itself is characterized by a combination of covalent (*i.e.*, Fe-$X$) and metallic (*i.e.*, Fe-Fe) bonding. We note that among these seven systems, the 122-system has arrested big deal of attention owing to the availability of high quality single crystals. To substitute for the Fe-site, transition metals, including Cr, Mn, Ru, Co, Rh, Ir, Ni, Pd, Pt, Cu, Zn, etc. are the choice.

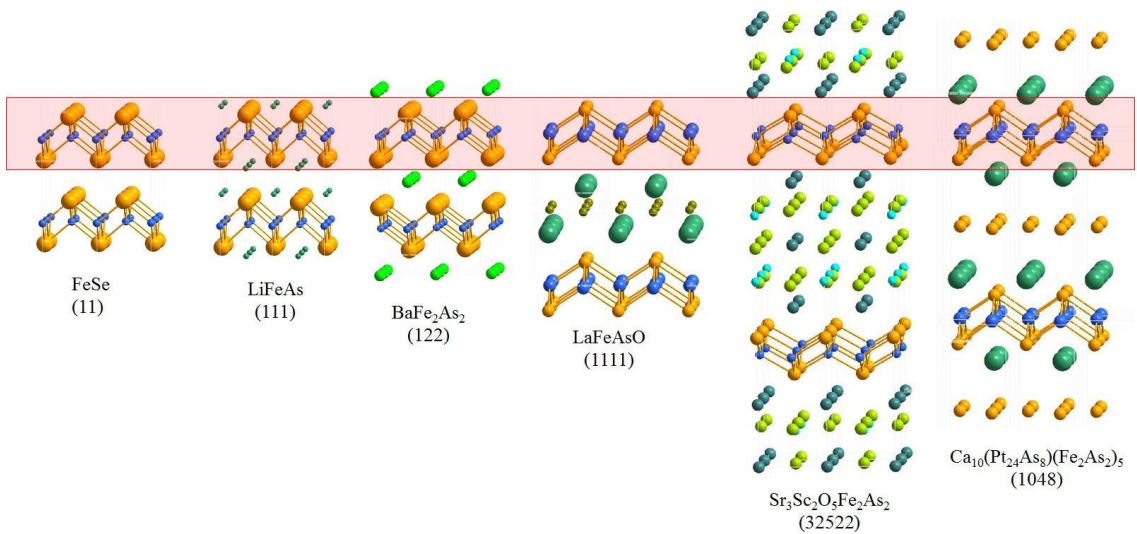

**Figure 5.** Crystal structures of the FBS, where the red shadow indicates the Fe$_2$$X$$_2$ ($X$ =As, P, Te, or Se) layers. All seven series of compounds have the tetragonal lattice structure with space-group of P4/nmm [1-8].

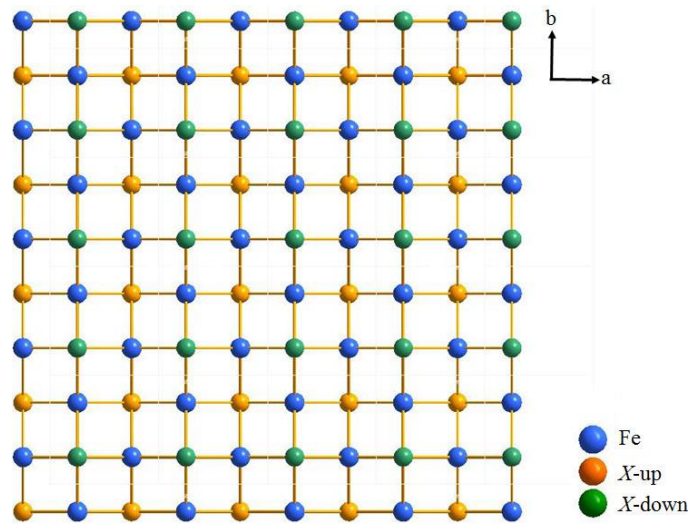

**Figure 6.** Structure of the Fe$_2$$X$$_2$ ($X$ =As, P, Te, or Se) layer, where half of the $X$ atoms lie above the Fe plane, while the others lie below the plane.

Generally, the FBS was synthesized through the solid-state reaction in an inert atmosphere or vacuum. Besides, the solid-state reaction can also be implemented under high-pressure as few tens of thousands atmosphere pressure. To distinguish these two methods, we name them as ambient-pressure and high-pressure techniques. It is worth noting that the growth of impurity-doped epitaxial thin film by pulsed-laser deposition is also a promising way, but there is no any report yet.

*(i) Ambient-pressure technique*

Up to now, most of our present FBS were synthesized under ambient-pressure which can be divided into several steps [1-3, 46, 47]: firstly, the starting materials are stoichiometrically mixed together. Secondly, the mixture was sealed into a tantalum capsule, which was sometimes with an *h*-BN inner to isolate the sample from the capsule, because of the toxicity and low vaporization temperature of arsenic/phosphorus elements and the high reactivity of rare-earth, alkali, and alkali-earth metals. Thirdly, the as-prepared pellets were sealed in an evacuated quartz tube and then heat-treated, where the heating conditions may vary from different material systems and different labs. Fortunately, apart from the oxygen-deficient system like $A$FeAsO$_{1-\delta}$, of which the sample quality may very sensitive to the synthesis procedure like the heat-treatment process and even the used furnace as those of oxygen-deficient cuprate, most of other Fe-based systems are quite reproducible. Here, we take the 122-type (Ba,K)Fe$_{2-2x}$Zn$_{2x}$As$_2$ as an example [46, 47]. The arsenide compounds as BaAs, KAs, FeAs, and ZnAs were synthesized from the metal pieces (Ba, K, Fe, Zn, *etc*.) and As powders at a relatively low temperature (500 - 700 ℃), which is essential to avoid the loss of As in the next steps and also to obtain fine powder of all starting materials other than metal pieces. The stoichiometric mixture of BaAs (lab made), KAs (lab made), FeAs (lab made), Fe (3N), and Zn (4N) or ZnAs (lab made) was placed in a tantalum capsule with an *h*-BN inner (preheated in advance at ~1900 ℃ for 1 hour in nitrogen). The loaded capsule was sealed in an evacuated quartz tube under vacuum condition. The sample was finally heated at 1000 ℃ for 72 hours, and then slowly cooled down to room temperature at a rate of 20 ℃/hr.

*(ii)High-pressure technique*

In a high-pressure process the as-prepared pellets were sealed in BN crucibles [48-50], then placed into a graphite heating tube and mounted into a high-pressure sintering equipment for the solid-reaction process, which can be belt-type or cubic-type with different kinds of anvils. Here, the pressure, target temperature, and heat-treatment time all strongly depend on the material system and doping type. For the 1111-system [48], the pressure should be up to 6 GPa, while less to 3 GPa for the 122-system. We take 122-system as an example [49, 50], e.g., $BaFe_{2-2x-2y}Zn_{2x}Co_{2y}As_2$ and $Ba_{1-y}K_yFe_{2-2x}M_{2x}As_2$ ($M$ = Mn, Ru, Co, Ni, Cu and Zn). The stoichiometric mixture and capsule sealing processes resemble those of ambient-pressure technique, but here 50 wt.% of BaAs for $BaFe_{2-2x-2y}Zn_{2x}Co_{2y}As_2$ and 50 wt.% of BaAs/KAs for $Ba_{1-y}K_yFe_{2-2x}M_{2x}As_2$ were added as flux for single crystals growth. The capsule was compressed at 3 GPa in the belt-type high-pressure apparatus and heated at 1300 ℃ for 4 hours, followed by a decrease to a constant temperature of 1100 ℃ for 1 hour. Note that the pellet was self-separated into sizes of around $0.3 \times 0.2 \times 0.1$ mm$^3$ or much smaller after it left in vacuum for 2-3 days. Since additional BaAs and KAs were added in excess to be the flux, which can be washed away using pure ethanol. The single crystals can be cleaved along the $c$-axis.

## 3.2. Irradiation

For the heavy impurities doping level, the chemical substitutions may lead to inhomogeneity in the lattice sites, resulting in change of local electronic density and even the Fermi-surface topology, which may mask the intrinsic impurity effects on the superconductivity. Despite that Zn is similar as that of Cu in the cuprate superconductor system $YBa_2Cu_3O_8$, previous experiments demonstrated that Zn impurities cannot be substituted beyond 5% per Cu [20, 27], or some second phases like $Y_2BaCuO_5$ will appear and some Zn ions will gather around the surface of grains, which will mislead the nature of SC suppression effects caused by Zn. Alternatively, particle irradiation, generally used to introduce intrinsic defects or carrier artificially for the semiconductors, is a promising method to study the scattering centers.

The particles should be light elements as electron, proton, neutron, and low power $\alpha$-particle. However, most particle irradiation introduce correlated disorder and external defects as columnar and/or clusters, especially for the heavy particles like proton. Although the MeV-range electron irradiation can produce vacancy-interstitial (Frenkel) pairs as scattering centers [20], one should be extremely careful with identifying the defect types before concluding their nature of pair-breaking effects. **Figure 7** demonstrates various particle irradiation and the corresponding defects types [87]. The profiles of the irradiation induced defects strongly depend on the irradiation particles and power. To create uniformly distributed point defects over the entire crystal, it is essential to apply long attenuation length and the small recoil energy, such as electron irradiation with the small mass of electron particles. Nevertheless, the relatively large recoil energy may induce complex defects such as clusters and cascades, and even bulk defects as columnar tracks. Besides, the sample itself is also crucial for the irradiation experiments, for which the crystal structure, composition, and even the sample geometry can influence the irradiation results. Therefore, the particle irradiation can hardly replace the chemical doping completely.

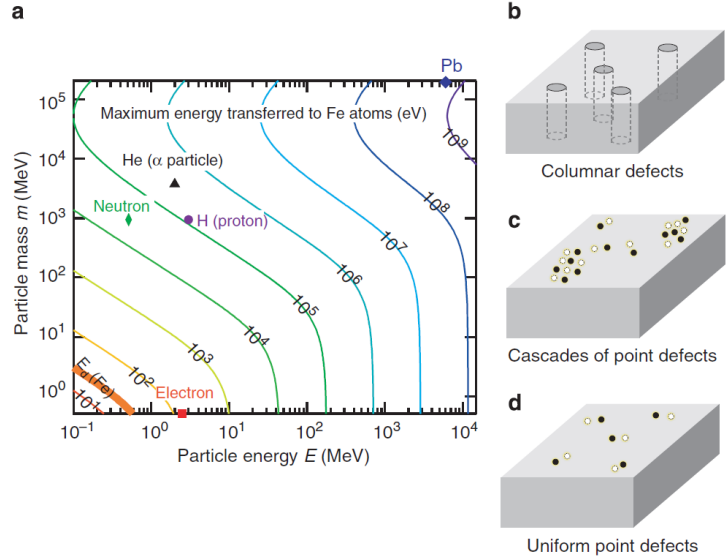

**Figure 7**. Various particle irradiation and the corresponding defects types. (a) Contour plot of particle mass $m$ and particle energy $E$ for various irradiations on the maximum energy transferred to Fe atoms. Typical particles of electron, neutron, and proton are marked as red square, green diamond and purple circle, respectively. Schematic images for three types of irradiation induced defects, including (b) the columnar defects, (c) cascades of point defects and (d) uniform point defects, where the last defects were expected as the ideal ones from the electron irradiation, from Mizukami *et al.* [87].

## 4. Experiments on impurity effects

### 4.1. Controlled impurities within the $Fe_2X_2$ planes

Up to now, most of the previous impurity-doping studies have been carried out on the polycrystalline samples synthesized under ambient-pressure. **Figure 8** shows a scanning electron microscopy image for the Zn-doped $Ba_{0.5}K_{0.5}Fe_{1.9}Zn_{0.1}As_2$ polycrystal from Cheng *et al.* [51]. The crystal was detected by energy dispersive x-ray (EDX) in several points, and the compositions of Zn in different grains were found as close to the nominal one. However, one possibility is that the Zn may filter onto the surface of grains, which cannot be detected by EDX because the technique cannot eliminate the surface element from the real concentration of grains.

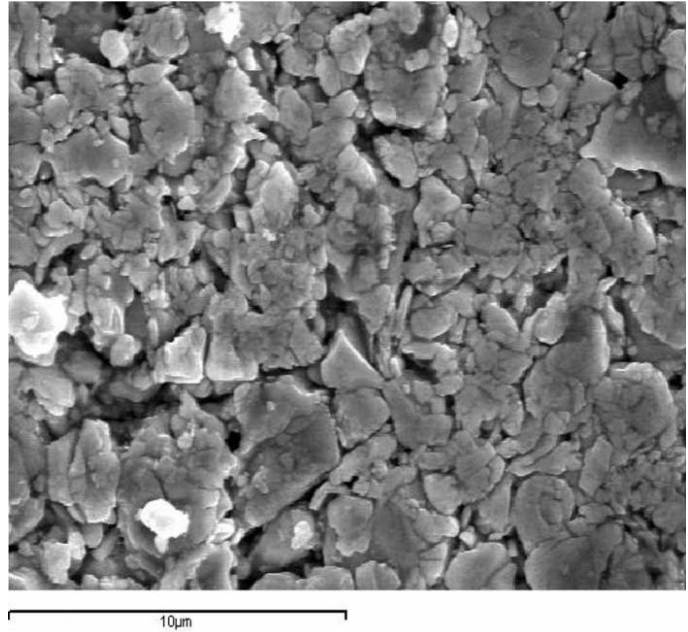

**Figure 8**. Scanning electron microscopy image of Zn-doped $Ba_{0.5}K_{0.5}Fe_{1.9}Zn_{0.1}As_2$ polycrystal. Several grains with diameters of ~ 1 $\mu$m were chosen randomly to carry out the measurement of energy dispersive x-ray. The EDX data suggested the compositions of Zn were accordance with that of normal content, from Cheng *et al*. [51].

To avoid the possibility of that Zn can precipitate, high-quality single crystals are essential for the study. Compared with the polycrystal, a single crystal often has substantially lower defects and less strain [20]. A single crystal also has sizable and atomically ordered surfaces, which are vital for surface sensitive measurements, such as STM [45, 52, 53] and ARPES [54-56]. As we mentioned in **Section 3.1**, most of single crystals are from 122-system, owing to the best quality and stability [58]. However, it is still extremely difficult to grow the high-doping level and high-quality Zn-doped Fe-based single crystals, for which the main difficulty is from the low melting point (419 ℃) of Zn which is far below the reaction temperature (around 800 - 1000 ℃).

On the other hand, although one can select ZnAs alloy instead of Zn metal [51], such method has been scarcely used due to the difficulty of ZnAs preparation.

Tan *et al*. [58] utilized the starting material of $Fe_{2-x}Zn_xSe_2$ instead of Zn or ZnSe, and synthesized the hole-type single crystal $K_{0.8}Fe_{2-y-x}Zn_xSe_2$, as shown inside **Figure 9**. The XRD and EDX results demonstrated that the Zn had been successfully doped into the crystal. This is an impressive method for the future growth of other single crystals under ambient-pressure.

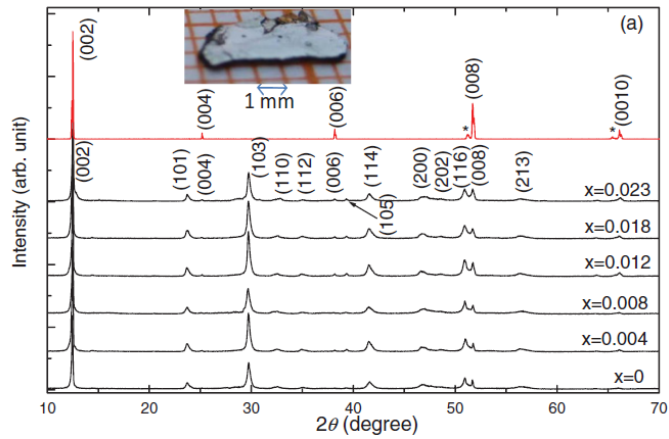

**Figure 9**. Powder x-ray diffraction patterns for the $K_{0.8}Fe_{2-y-x}Zn_xSe_2$ single crystals with *x*=0-0.023. The red curve is the single-crystal XRD pattern of a $K_{0.8}Fe_{2-y}Se_2$. An optical photo of the as-grown single crystal is given on the top of insert, from Tan *et al*. [58].

Under the ambient-pressure conditions, it is generally necessary to synthesize the single crystals for long time, for instance, 92 hours for $K_{0.8}Fe_{2-y-x}Zn_xSe_2$ [58]. To shorten the time, the high-pressure method is a promising approach to synthesize Zn-doped single crystals, owing to the virtues of high reaction temperature and short time synthesis period typically of only a few hours. **Figure 10** shows typical single crystals of Zn-doped $Ba_{0.5}K_{0.5}Fe_{1.95}Zn_{0.05}As_2$ grown by high-pressure method. The crystals crystallize into the tetragonal $ThCr_2Si_2$-type structure (*I4/mmm*), which has a two-dimensional feature. However, the two-dimensional anisotropic factor is not as large as those of cuprates or other 2-D materials, and few reports have been presented on the cleavable of the crystals, except for some studies by using special techniques like STM and ARPES. The high-pressure grown single crystals can be cleaved by using scotch tape. Indeed, cleaving technique is indispensable to explore the real concentration of Zn in the crystal, and consequently to elucidate the intrinsic properties of Zn substitution effect on SC. In **Figure 10(c),** a flake-like crystal was cleaved on a scotch tape. The crystal demonstrates mirror-like surface and free of inhomogeneous distribution or grain boundary. Such thin crystals were held on silicon substrate using epoxy as shown in **Figure 10(a) and (b)**, which were then used for the electron probe micro-analyzer (EPMA), EDX, x-ray diffraction, and transport properties measurements.

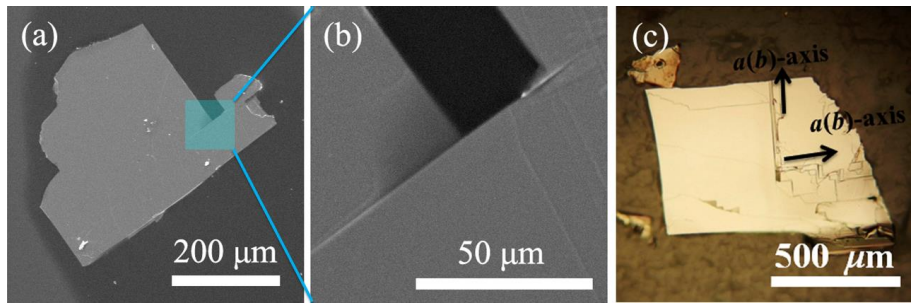

**Figure 10.** (a) and (b) SEM and (c) optical images of the $Ba_{0.5}K_{0.5}Fe_{1.95}Zn_{0.05}As_2$ single crystals. The crystals were firstly cleaved along *c*-axis into a clear surface as shown in (c), where we can see the cleaved crystal on tape, and some perpendicular steps corresponding to the *a*- or *b*-axis, owing to the tetragonal structure. The cleaved thin crystals were held on a silicon substrate by epoxy, and cleaved again before deposited a 20 nm gold layer for the SEM, EDX, and some other measurements.

EPMA and EDX measurements on such cleaved single crystals showed that the real concentration of Zn is in well accordance with the nominal one. Powder and single crystal x-ray diffraction measurements demonstrated (see **Figure 11**) that the impurities induced systematic change of lattice parameters. Thus, the expected difficulties of the substitution due to the high volatility of Zn and the electronic difference between $Zn^{2+}$ and $Fe^{2+}$ had been really overcome by using high-pressure technique.

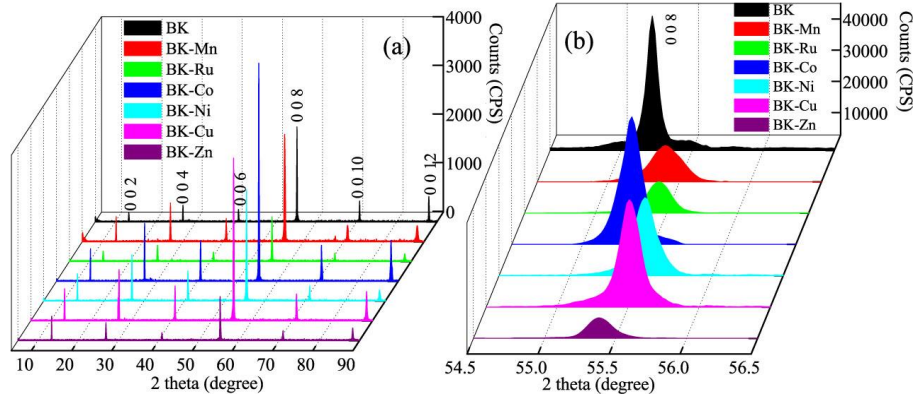

**Figure 11**. XRD patterns of the single crystals of $Ba_{0.5}K_{0.5}Fe_{2-2x}M_{2x}As_2$ doped with various impurities of $M=$ Mn, Ru, Co, Ni, Cu and Zn. Here the impurity concentration $x=0.05$ is as the nominal value, from Li e*t al*. [49].

## 4.2. Single impurity in superconducting state

Single NMI dot $Zn^{2+}$ has a filled $d$-shell, and hence works as quasiparticle interference scattering and provides information of the gap structure of the superconductors [1-4]. Since any sort of impurity or defect in a crystal can be screened by the conducting electrons which leads to the well-known Friedel oscillations of the charge and spin density around the imperfection. In real space, interference among such oscillations stemming from random impurities is currently irresolvable in these systems, but the Fourier transform of the measured electron density will reflect the structure of the charge susceptibility in reciprocal space. A natural technique to map out the electron density near the Fermi level is by scanning tunneling spectroscopy (STM) which worked well for the study on cuprates, especially on $Bi_2Sr_2CaCu_2O_8$ [20, 52, 53]. It is worth noting that the most crucial requirements in this experiment are doping impurity into the SC layer and such crystal should be in high quality free of other defects. In addition, the single crystal should behave as 2-D structure as to be mechanically cleaved to obtain atomically flat and clean surfaces.

Zhu and co-workers [59] have calculated the effects of a single NMI on the $(K,Tl)Fe_xSe_2$ superconductors, within both a two-band model and five-band model. **Figure 12** gives the estimated density of states in the system without impurity and the local density of states at the site nearest neighboring to the NMI dot. The authors found that the impurity-induced resonance state can only exist for a $d_{x^2-y^2}$-wave pairing state. In addition, they also found that the bound-state peak in the local density of states occurs at a nonzero energy even in the unitary limit, indicating an opposite situation from the cuprate systems. The prediction arouses subsequent various theoretical studies [60-64]. Chen *et al*. [60] calculated the local effect of Zn on the 122-type $Ba(Fe_{1-x-y}Co_yZn_x)_2As_2$, and suggested quite short correlation lengths $\xi_\Delta$ for the case of $s_\pm$ wave state, which is only two lattice sites, namely, $\xi_\Delta$ is about two lattice distance of Fe ions. Such a short screening length was attributed mainly to the strong local Coulomb repulsion $U$ which acts on the charge sector.

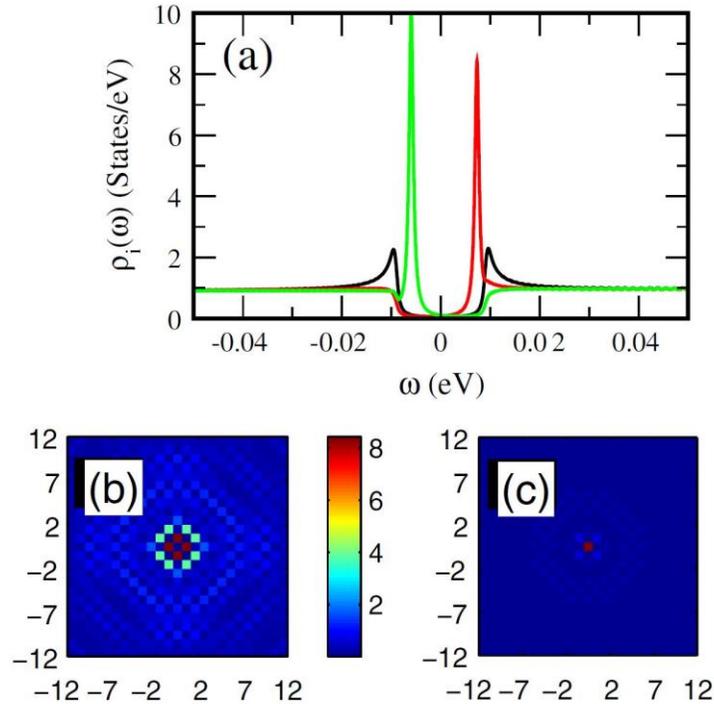

**Figure 12**. Simulation results for the local density of states of a single NMI site. (a) The local density of states at the first nearest neighboring site, where the black line corresponds to the bare density of states, red line to the local density of states without interband scattering ($u = 100$ eV and $v = 0$), and green line with black line ($u = v = 100$ eV). (b), (c) The local density of states imaging in the superconductor with ($d1$) pairing symmetry without and with interband scattering, respectively, from Zhu e*t al*. [59].

However, for the STM measurement on the Zn-doped single crystals, there is no any report on FBSs up to now. The main reason lies in the technique difficulty in growth of high quality single crystals, as discussed in **Section 4.1**. Moreover, to probe the local effect of Zn ion the surface layer should be Fe-$X$ plane, but not the barrier layer, i.e., the $A$-site ions as Ba, Sr, Ca, K, Eu, *etc*. However, the terminal ions are often end up on either of the two cleaved Fe-$X$ surfaces, masking the profile of Zn. Previous successful observations of atomically resolved STM images are mostly on $A$-ions terminal, few reports on the As-ion terminal, depending on the materials system and cleaving temperatures [45].

Recently, Yang and co-workers [65] studied the scattering of Cu impurities on Na(Fe$_{0.97-x}$Co$_{0.03}$Cu$_x$)As via the STM, instead of Zn. They considered the Cu as a NMI or weak magnetic impurity comparing with the strong magnetic host ions Fe$^{2+}$. **Figure 13** shows spatial maps of local density of states measured at different energies. The local density of states around the Cu impurity exhibits a systematic evolution with a spatial length of about 1.5-2.0 nm being about 5-7 Fe sites, which is well consistent with the coherence length. Therefore, the Cu impurity was believed to result in a point disorder, leading to decay of the in-gap quasiparticle states in a scale of the coherence length. Consequently, Cu impurities induce Cooper pair-breaking in the strongly anisotropic $s_{\pm}$ pairing symmetry state.

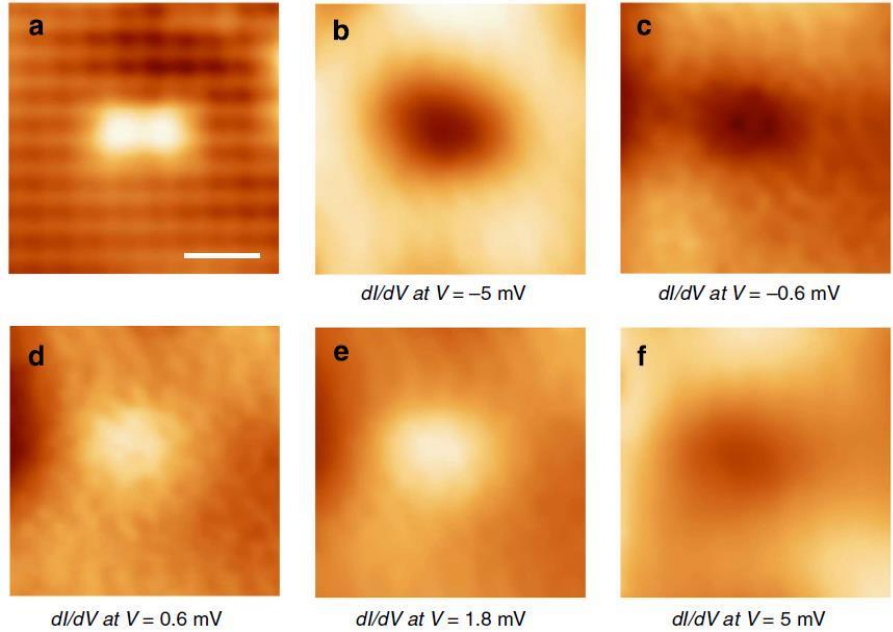

**Figure 13**. Local density of states mapping of the first nearest neighboring site of Cu cite in Na(Fe$_{0.97-x}$Co$_{0.03}$Cu$_x$)As superconductors. (a) The topographic image of a Cu impurity, where the scale bar is 1 nm. (b-f) The mapping of the local density of states measured at bias voltages of -5, -0.6, 0.6, 1.8 and 5mV, respectively. The spatial influence of a Cu impurity is about 20Å, being about 5-7 Fe sites as the corresponding coherence length, from Yang e*t al*. [65].

To explain the reduction of superconducting regions by Zn impurity in the cuprate superconductor, Nachumi *et al*. [66] proposed a 2-dimensional "swiss cheese" model based on the muon spin relaxation measurements. In this model, the charge carriers are excluded from superconductivity around each Zn ion, resulting in non-superconducting regions with the diameter of characterized coherence length $\xi_\Delta$. The result was consistent with the subsequent magnetic susceptibility measurement [67].

In the FBS, a most recent work demonstrates the local destruction of superconductivity by nonmagnetic Zn impurities in Ba$_{0.5}$K$_{0.5}$Fe$_2$As$_2$ by exploring phase-slip phenomena in a mesoscopic structure with 119×102 nm$^2$ cross-section [68]. However, the impurity-free nanobridges demonstrated thermal stability due to smaller number of defects. Considering the general condition for the appearance of phase-slip, the cross-sectional area should be comparable with the value of $\sqrt{2}\pi\xi$ [69-71]. However, the $\xi_{ab}$ and $\xi_c$ of $Ba_{0.5}K_{0.5}Fe_2As_2$ were estimated to be only 2.05 and 1.20 nm, respectively, two orders magnitude less than the cross-section of the nanowire. The Zn impurity was therefore proposed to suppress superconductivity in a "Swiss cheese"-like pattern as that of cuprate superconductors. Since the 122-type superconductors possess weakly anisotropic layered structure, the Cooper pairs reside both in and out-of the $Fe_2As_2$ superconducting planes, indicating a 3-D "Swiss cheese" model, where the order parameter can fluctuate along abundant narrow superconducting channels with both in-plane and out-of-plane directions as shown in **Figure 14**. For a conventional superconducting gap like $s_{++}$, the nonmagnetic impurity ions work as point defects, but do not affect the Cooper pairs. As a contrast, the Zn ions can induce local destruction of superconductivity for the unconventional $s_{\pm}$ pairing symmetry, and consequently result in phase-slip phenomenon in the $Ba_{0.5}K_{0.5}Fe_{1.94}Zn_{0.06}As_2$ nanobridges. The local destruction may provide an evidence for the pair-breaking effect of nonmagnetic impurities and the unconventional $s_{\pm}$ pairing symmetry for the iron pnictide superconductors.

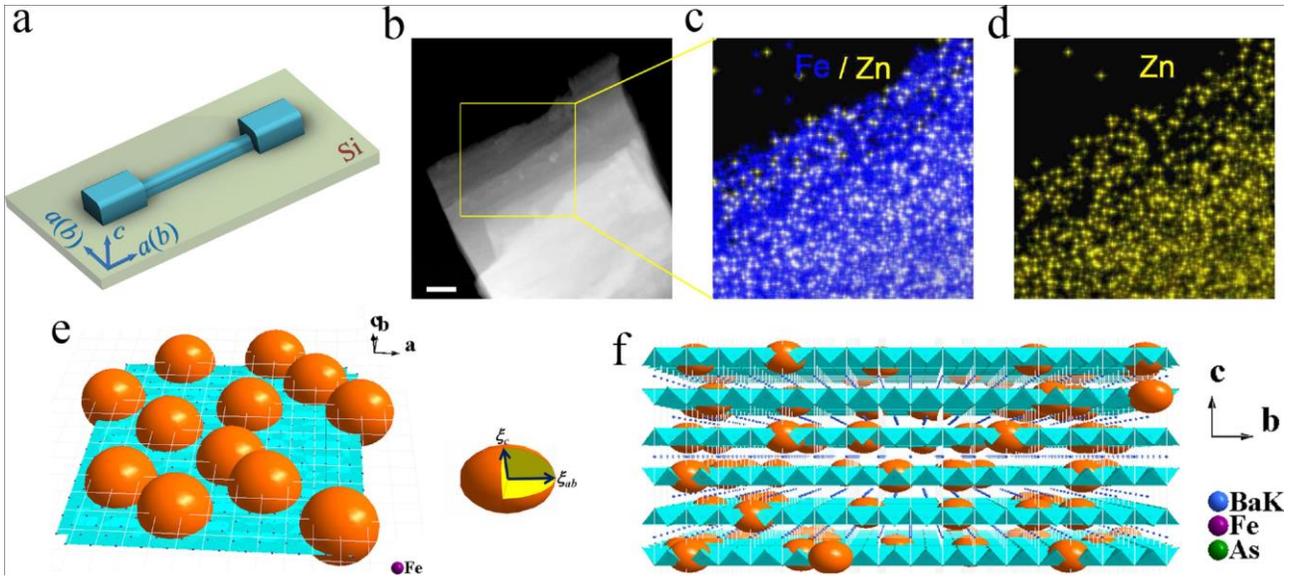

**Figure 14** (a) Transport measurement scheme along a nanobridge. The current flows along the *ab*-plane. (b) High angle annular dark field scanning transmission electron microscopy image of a Ba$_{0.5}$K$_{0.5}$Fe$_{1.94}$Zn$_{0.06}$As$_2$ flake, where the crystal was detected along the *c*-axis. The scale bar represents 50 nm. (c) and (d) Scanning transmission electron microscopy energy dispersive X-ray spectroscopy mapping for Zn/Fe and Zn distributions, respectively, within the area indicated in (b). (e) and (f) Schematic representation of the 2-D and 3-D "Swiss cheese" models, respectively. The yellow oblate spheroid corresponds to the non-superconducting regions centered on Zn ions with an equatorial length $\xi_{ab}$ and a polar length $\xi_c$, from Li *et al*. [68].

## 4.3. Magnetic properties

Since the SC mechanism for the FBS was widely considered as a spin fluctuation origin, magnetic properties are of great importance to understand the NMI effects on the SC [5-12]. First of all, magnetic susceptibility measurement provides a direct determination of $T_c$ owing to the Messier effect. More importantly, since the FBS demonstrates metallic nature for the parent compounds before the carrier doping, and an antiferromagnetic (AF) order phase appears below the critical temperature of spin density wave (SDW) ordering at $T_N \approx 130$ K, as well as in the under-doped regime, substitution of spinless Zn$^{2+}$ ions will be greatly interesting to explore information of the pair symmetry and even the SC mechanism [72-77]. However, compared with the transport properties, magnetic properties of the Zn-doped FBS are reported in very few works. In this section, we will briefly summarize some present work of the Zn effects on AF state and local magnetic moments.

### *(i)Antiferromagnetic state*

In the SDW state, the ordered magnetic moment is about 0.3 $\mu_B$ and the ordering vector is Q ≈ ($\pi$, 0). SC was induced by suppression of $T_N$ via carrier doping, the AF phase however generally exists within the underdoped regime. It was expected from the single-defect calculation [73-77] that the substitution of spinless $Zn^{2+}$ ($s = 0$) ions for strongly magnetic host ions of $Fe^{2+}$ ($s = 2$) may result in a strong local suppression of short-range AF order, hence the Zn may also suppress $T_N$ as what was done from the carrier doping.

Li and co-workers studied the Zn-doping effect on the polyscrystalline LaFeAsO parent compound [78]. **Figure 15** shows the temperature dependence of magnetic susceptibility ($\chi$) of LaFe$_{1-x}$Zn$_x$AsO with various doping level of Zn. The insets show the enlarged plots for $x = 0$ and 0.02. The arrows indicate the anomaly in susceptibility related to structural and/or magnetic transitions. Slight Zn doping in LaFe$_{1-x}$Zn$_x$AsO drastically suppresses the AF in the parent compound. For instance, the $T_N$ is decreased to 137 K with 2at.% doping of Zn, slightly less than that of Zn-free sample (150 K). However, the SDW can hardly be observed with Zn up to 5at.%. Since the sample is polycrystalline and the SDW order is weak even for the impurity-free sample, the sample quality may dominate the SDW other than the Zn-doping. Even though, the magnetic data together with their resistivity properties indicated that the SDW order was sensitive to Zn doping. Zhang and Singh [77] proposed two mechanisms for the $T_N$ suppression from Zn doping: (i) Zn ion provides localized states and disrupts the electronic structure of the Fe sheets near Fermi level. (ii) It introduces local moments that have the pattern of the checkerboard antiferromagnetic state mainly on the four neighboring Fe.

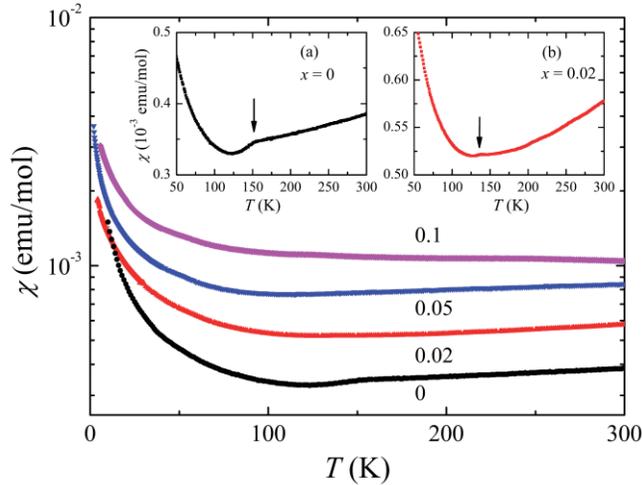

**Figure 15.** Magnetic susceptibility ($\chi$) *vs.* temperature for Zn-doped parent compound LaFe$_{1-x}$Zn$_x$AsO. The insets show the enlarged plots for samples with $x = 0$ and $0.02$. The arrows indicate the anomaly in susceptibility related to structural and/or magnetic transitions, from Li *et al*. [78].

### (ii) Local magnetic moments

The impurity ions of Zn$^{2+}$ work as spinless centers, which may induce moments of $s = 2$ on the Fe-sites. However, there is still an open question for the Zn-induced local moment behavior and no direct experimental evidences can support it. For instance [62-64, 79, 80], the first question is whether a nonmagnetic site can induce a free paramagnetic moment in a metallic correlated system. It should be noted that the study on Zn suppression effects on $T_N$ should be focused on the under-doped or undoped regimes where the AF state is still exist. A few candidates can be used to explore the Zn-induced local moment, such as STM, muon spin rotation, neutron scattering, specific heat, magnetic susceptibility, and so on. Among these methods STM measurement is the best and most direct method as we introduced in detail in **Section** 3. However, the STM analysis was plagued by materials problems as in **Sec 4.2**, as well as the muon spin rotation and neutron scattering measurements. Apart from the specific heat experiment as we will introduce in **Section 4.5**, magnetic susceptibility measurement is a promising and simple method.

Theoretically, once substitution of Zn provides the local spin moment, it will enhance a Curie-like behavior in the uniform magnetic susceptibility, despite that the host material represents the metallic conduction of carriers. Curie-like behavior for the characteristics of temperature dependent magnetic susceptibility $\chi(T)$ is given by [20]

$$\chi(T) = \chi_0(T) + \frac{NP_{eff}^2\mu_B^2}{3k_B(T-\Theta)} \tag{16},$$

where $\chi_0(T)$ corresponds to the $\chi(T)$ without Curie-Weiss background which depends on the carrier density of the impurity-free sample, $N$ is the number of magnetic ions, $\Theta$ is the Curie-Weiss temperature, and $P_{eff}$ is the effective moment in the units of the Bohr magnet ($\mu_B$) and relates to the Zn doping. Therefore, by fitting the temperature dependent magnetic susceptibility $\chi(T)$ for different doping level of Zn, one can systematically obtain the $P_{eff}$. Although an initial susceptibility data on Zn-substituted $Ba_{0.5}K_{0.5}Fe_{2-2x}Zn_{2x}As_2$ polycrystals did suggest yielding of moments from $0.482\mu_B$/Fe to $0.362\mu_B$/(Fe+Zn) [51], the nature of impurities contribution to the magnetic susceptibility can only been determined since, in carefully impurity-controlled samples as discussed in **Section 4.1**. Future comprehensively study on Curie-Weiss behavior of Zn-doped single crystals is essential.

In addition, nuclear magnetic resonance (NMR) can probe nuclei coupled to the SC $Fe_2X_2$ planes which is capable of yielding insights into the local magnetic structure. Using a variety of nuclei, NMR can develop a fairly clear picture of how a nominally NMI provides a cloud of local staggered polarization on nearby Fe-site, and can characterize the response as a function of doping and temperature, for different materials. Kitagawa and co-workers [79] studied the Zn-substituted LaFeAsO$_{0.85}$ polycrystal using [75]As and [139]La NMR and nuclear quadrupole resonance (NQR) (see **Figure 16**). Although SC in LaFeAsO$_{0.85}$ disappears by 3% Zn substitution as we will introduce in **Section 4.6.1** [48], it was found that NMR/NQR spectra and NMR physical quantities in the normal state can hardly been changed, indicating that the crystal structure, electronic states and magnetic moments are not modified by Zn substitution. The results suggest that the suppression of SC by Zn substitution is not due to the change of the normal-state properties, but the strong nonmagnetic pair-breaking effects.

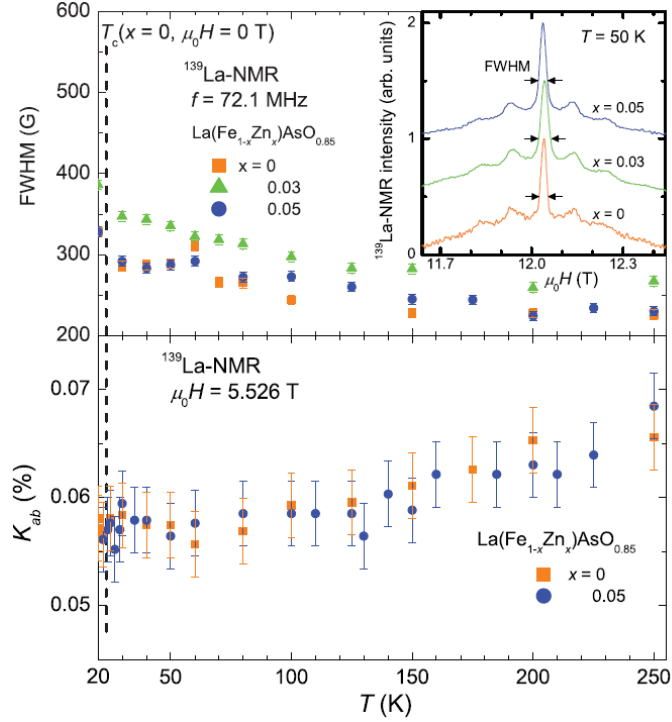

**Figure 16**. Temperature dependence of the full width at half maximum (FWHM) of $^{139}$La NMR spectra for polycrystalline LaFe$_{1-x}$Zn$_x$AsO$_{0.85}$ substituted with Zn of 0, 3%, and 5 %. The Zn was observed weak change for FWHM. (Inset) Field swept $^{139}$La NMR spectra at 50 K at 72.1 MHz. The lower panel gives temperature dependence of $^{139}$La NMR Knight shift for $H$ // $ab$, from Katagawa e$t$ $al$. [79].

## 4.4. Transport properties

Transport properties measurements can probe the impurities or defects effects on various SC properties, including the carrier, coupling between charges, spin degrees of freedom, and so on [20]. All of these parameters are correlated with the induced suppression of $T_c$, and more importantly with the pair-breaking from impurities. In this section, we will firstly discuss the in-plane transport properties in both high-temperature and low-temperature (around $T_c$) regions. Although transport properties of the 2-D FBS show slightly anisotropic along $c$-axis and $ab$-plane, we will focus on the in-plane properties where the impurities scattering occurs within the SC $Fe_2X_2$ planes.

### (i) In-plane transport properties

Studying on the normal state in-plane resistivity is the most common way to explore the impurities influence. Apart from few systems like hole-doped (e.g. $Ba_{1-x}K_xFe_2As_2$ and $K_{1-\delta}Fe_{2-y}Se_2$ [49, 81]), most of FBSs behave as a typically metallic nature, especially for the optimal-doped crystals. The in-plane resistivity ($\rho_{ab}$) decreases linearly with temperature above $T_c$ as

$$\rho_{ab} = \rho_0 + kT \tag{17},$$

where $\rho_0$ is the residual resistivity which will be discussed in detail in the **Section** 4.4.2.

According to the Mathiessen's rule [81], the impurity scattering rate in classical metals adds incoherently with the inelastic scattering rate in the pure material, usually owing to phonons. The normal state $\rho_{ab}(T)$ consequently will display upwards in parallel. The approximate additives of the impurity scattering was found for the $\rho_{ab}(T)$ of Zn-doped $BaFe_{1.89-2x}Zn_{2x}Co_{0.11}As_2$ ($x$ = 0-0.08) as shown in **Figure 17** [50]. However, the parallel shift phenomenon occurs at high temperature regions. For the low-temperature regions around $T_c$, a slight upturn was observed as increasing doping level. Such weak low-$T$ upturn behavior was also found in cuprate superconductors, which could be attributed to some factors as: (i) disorder, due to that the spinless impurity may result in a scattering on magnetic perturbations induced in the planes, as discussed in **Section** 2.2.2; (ii) charge localization, leading to a metal-insulator transition (MIT) with decreasing carrier content, which can been evidenced by the low-$T$ upturn of Hall coefficient.

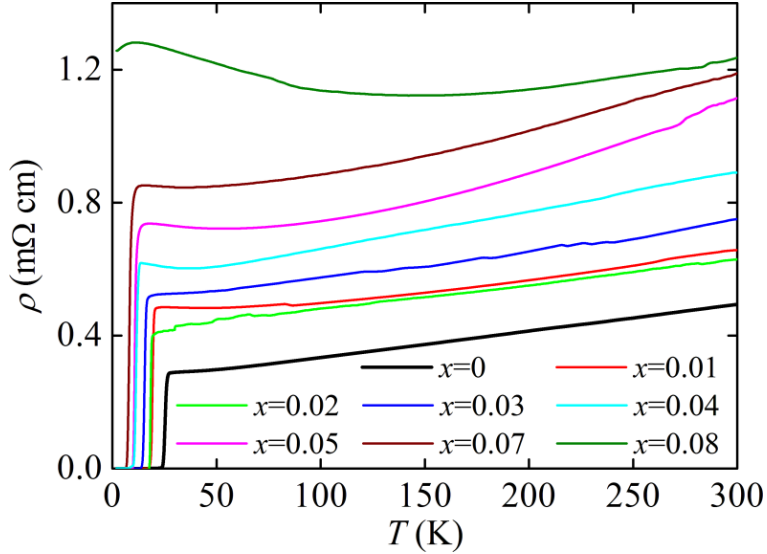

**Figure 17**. Temperature dependence of in-plane resistivity ($\rho_{ab}$) for the BaFe$_{1.89-2x}$Zn$_{2x}$Co$_{0.11}$As$_2$ with Zn content of $x$ = 0-0.08, from Li e*t al*. [50].

Since the impurities suppression on $T_c$ is an additional element of characterization available although it is not solely linked to the properties of the normal metallic state, as discussed in **Section 3.2**, the transport properties measurement strongly depends on the crystal quality and the measurement technique. Otherwise, the impurity concentration, fractional site occupancies, and actual carrier density of single crystals cannot always be accurately determined. Particle irradiation is one of the promising method, as one can span a series of defect concentrations using only one single crystal by increasing the irradiation dose progressively. **Figure 18** shows temperature dependence of $\rho$ for the NdFeAsO$_{0.7}$F$_{0.3}$ single crystal under $\alpha$-particle irradiation in different steps [82]. At high $T$ far above $T_c$, the $\rho(T)$ curves are observed as upward parallel shift, indicating the enhancement of the hole content introduced by irradiation. On the other hand, a much more obvious upturn than that of chemical doping appeared in the low-$T$ regions, which was interpreted in terms of a MIT, the so-called Kondo-like phenomenon. Similar results were also found for the neutron irradiated polycrystalline LaFeAsO$_{0.9}$F$_{0.1}$ [83]. However, such low-$T$ upturn behavior was neither seen for the proton irradiated Ba(Fe$_{1-x}$Co$_x$)$_2$As$_2$ single crystals [84, 85], nor electron-irradiated Ba$_{1-x}$K$_x$Fe$_2$As$_2$ [86], BaFeAs$_{2-x}$P$_x$ [87] and BaFe$_{1.76}$Ru$_{0.24}$As$_2$ [88]. This is probably because the defect profiles are different from those of $\alpha$-particle and proton irradiation induced samples, for which the $\alpha$-particle irradiation defects were proposed as both magnetic and nonmagnetic scatterings, while the light particles of proton and electron were suggested to induce nonmagnetic ones. The scattering behavior of these experiments will be discussed in details in **Section 5**.

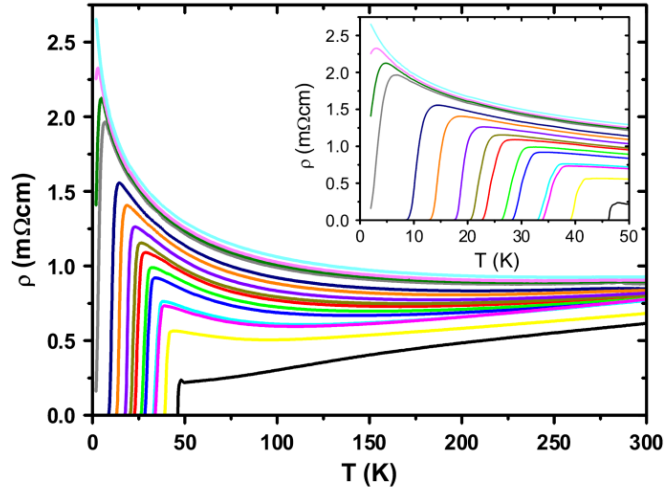

**Figure 18**. Temperature dependence of resistivity for the NdFeAsO$_{0.7}$F$_{0.3}$ single crystal under $\alpha$-particle irradiation in different steps, where the black curve corresponds to the pristine sample and the curves from yellow to cyan indicates the data after each step of irradiation. Inset shows magnified temperature region near $T_c$, from Tarantini e*t al.* [82].

#### *(ii) Improvement on measurement method*

The FBS are semimetallic at the normal state, and the resistivity is generally in unit of $\mu\Omega$ cm. Thus, it is rather challenged to measure the transport properties for a bulk crystal with a millimeter-size. In the traditional four-probe measurement technique, a high current bias is often necessary to enhance the measurement resolution, especially for the Hall signal, owing to the large geometry and strongly metallic behavior of the crystals. In this case, the heat effects can be a serious problem, especially at the interface between silver paste and crystal. To improve the experimental accuracy, micro- and nano-patterning techniques are promising ways to enhance the resistance and Hall signal.

Recently, we developed a micro-patterning technique to fabricate the microbridges on the 122-type FBS [89, 90]. The crystal was firstly cleaved into pieces with 1-2 $\mu$m thickness using scotch tape as shown from **Figure 10**, and then glued on Si substrate with the *ab*-plane parallel to the substrate surface using a thin layer of epoxy. The as-prepared crystal was then milled to a layer of 10 nm thickness by low energy argon ion beam (ion acceleration voltage of 200 V and beam current density $J_{Ar^+}$ of ~0.25 mA/cm$^2$), and immediately covered by a 100 nm layer of Au. To improve the interface contact between gold and crystal, we annealed the sample at 300 ℃ for 24 hours under nitrogen atmosphere, which can decrease the interface contact resistance to less than 0.1 $\Omega$ at room temperature. The crystals were fabricated as micro-devices as follows: (i) making microbridge patterns using the photolithography technique; (ii) argon ion milling the sample into a thickness of 300 nm; (iii) removing the photoresist by acetone and connecting the electrodes with silver epoxy; and (iv) etching the whole device until the crystal was completely removed except the micro-bridge. Note that the whole sample was etched in step (iv), except the parts under the silver paste.

**Figure 19** shows a scanning electron microscopic image of a Ba$_{0.5}$K$_{0.5}$Fe$_2$As$_2$ microbridge for both in-plane resistivity and Hall resistance measurements. We emphasize that, in the present method, high quality area of the crystal was carefully selected as for the microbridge with small geometry of $10 \times 4 \times 0.2$ $\mu$m$^3$. Besides, the electrodes are SC as well, so that we can restrict the heat effects considerably. Consequently, the measurements should be more actuate than those of bulk crystals. To measure the Hall coefficient the sample should be rotated along the axis of transverse current ($\vec{I}_{xx}$) to avoid the Lorentz force along the $\vec{I}_{xx}$, namely, $\cos(\vec{H} \cdot \vec{I}_{xx}) \equiv 0$. The sign of magnetic field can be changed via rotating the microbridge during the Hall coefficient measurements, where the effective magnetic field can be considered as $H = H_0 \sin(\vec{H} \times \vec{I}_{xy})$ and $\vec{I}_{xy}$ is the transversal current, i.e. the Hall current. To avoid the magnetoresistance for the transversal electrodes, we calculated the $R_{xy}$ by changing the sign of fields as

$$R_{xy}(H) = \left[R_{xy}(H^+) - R_{xy}(H^-)\right]/2 \qquad (18)$$

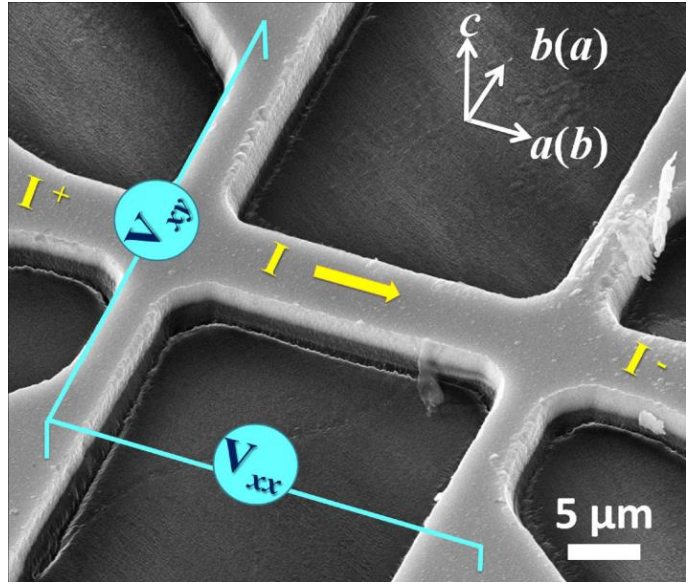

**Figure 19**. Scanning electron microscopic image of a microbridge for transport measurements, from Li e*t al*. [90].

For the present measurement setup, the resistance of microbridge is relatively large enough, normally above 20 $\Omega$, thus we can apply such a low current as 10 $\mu$A. **Figure 20** shows the temperature dependence of in-plane resistivity $\rho_{xx}$ for $Ba_{0.5}K_{0.5}Fe_2As_2$ (BK), nonmagnetic impurity Zn-doped $Ba_{0.5}K_{0.5}Fe_{1.95}Zn_{0.05}As_2$ (BKZn), and magnetic impurity Co-doped $Ba_{0.5}K_{0.5}Fe_{1.95}Co_{0.05}As_2$ (BKCo). The values of $\rho_{xx}$ were found as to be about one order magnitude less than our previous results on bulk crystals [49], suggesting the release of scattering from impurities or other imperfections, for which we can attribute to the improvement of crystal synthesis technique and measurement setup.

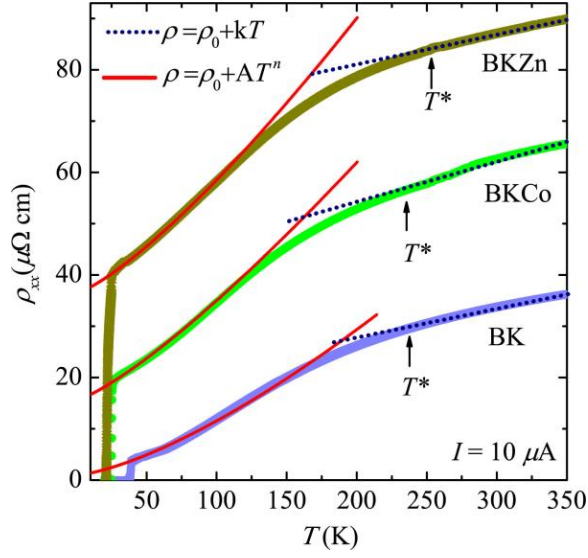

**Figure 20**. Temperature dependence of transverse resistivity for the $Ba_{0.5}K_{0.5}Fe_2As_2$ (BK), $Ba_{0.5}K_{0.5}Fe_{1.95}Zn_{0.05}As_2$ (BKZn), and magnetic impurity Co-doped $Ba_{0.5}K_{0.5}Fe_{1.95}Co_{0.05}As_2$ (BKCo). The blue lines indicate $T$-linear fitting for $\rho$ as $\rho = \rho_0 + kT$, and the red lines correspond to the $T$-nonlinear $\rho$ fitting as $\rho = \rho_0 + AT^n$, from Li *et al*. [90].

### (iii) Residual resistivity

Let us come back to **Eq**. (17), in which the $\rho_0$ should access zero for an ideal metal, while the appearance of non-zero $\rho_0$ is related to the scattering of carriers by the impurity atoms. However, SC of the high-$T_c$ superconductors were introduced by electron or hole doping, regardless of cuprate or FBS, which is hard to be free of impurity atoms and hence non-zero $\rho_0$. The behavior of $\rho_{ab}(T)$ slop is generally depended on various factors [20], including the host crystal, samples quality, and doping type. In this section, we will not discuss the scattering originated from the superconductor itself, but for the external scattering from impurity, namely, the impurity induced change of residual resistivity as $\Delta\rho_0$.

To estimate the $\rho_0$ owing to the impurities scattering, one can extrapolate the $\rho_{ab}(T)$ to 0 K, or by measuring the parallel shift of the $\rho_{ab}(T)$ curves at high-$T$ as what was done on the cuprate system [20, 91, 92]. Indeed, it is easy to evaluate from a metallic compound owing to the linear behavior of the $\rho_{ab}(T)$ slops, while difficult for the non-linear $\rho_{ab}(T)$ slops in some materials like hole-doped Ba$_{1-x}$K$_x$Fe$_2$As$_2$ and K$_{1-\delta}$Fe$_{2-y}$Se$_2$ [49,81]. In **Figure 20** one can find that the $\rho_{ab}(T)$ slops of Ba$_{0.5}$K$_{0.5}$Fe$_2$As$_2$, Ba$_{0.5}$K$_{0.5}$Fe$_{1.95}$Zn$_{0.05}$As$_2$ and Ba$_{0.5}$K$_{0.5}$Fe$_{1.95}$Co$_{0.05}$As$_2$ consist of a high-$T$ linear and low-$T$ nonlinear regions. The former region establishes a metallic-like resistivity dominated by phonon. The $T$-nonlinear regime can be fit using a power-law relation

$$\rho = \rho_0 + AT^n \tag{19},$$

where $\rho_0$ is the residual resistivity at 0 K. Surprisingly, all powers are considerably less than that of a normal Fermi liquid ($n = 2$) which is dominated by strong electron-electron interactions, probably suggesting the effect from critical antiferromagnetic fluctuations. However, it is impossible to conclude the quantum critical profile from the extrapolation of normal state $\rho(T)$ to low-$T$ since all fittings were based on the normal state results, unless the superconductivity can be suppressed by high magnetic field. Particularly, the $\rho_0$'s of both impurity-free and impurity-doped crystals are dramatically less than previous measurements on the bulk crystals. Again, the relatively low $\rho_0$ indicates the improvement of measurement setup.

On the other hand, the strongly low-$T$ upturn in $\rho_{ab}(T)$ makes it difficulty in estimation of the $\Delta\rho_0$ using linear or nonlinear fitting as shown in **Figure 18**. An alternative approach is to measure the parallel shift of the $\rho_{ab}(T)$ curves at low temperature region as described in [93]. In contrast, Nakajima and co-workers found no upturn in $\rho_{ab}(T)$ curve of BaFe$_{2-2x}$Co$_{2x}$As$_2$ single crystals irradiated by proton, although the SC was depressed as well. The defects induced by the proton irradiation were suggested as nonmagnetic scattering centers, while $\alpha$-particle irradiated defects should be complex which are not only the NMI but also others that may adjust the effective carrier density.

The residual resistivity can hardly provide accurate determinations of the scattering rate directly; instead, it is a promising method to seek information from the depression of $T_c$ induced by the scattering centers. The decrease of $T_c$ ($\Delta T_c$) can be related with $\Delta\rho_0$ as [20, 91]

$$\Delta T_c = -\frac{\pi \hbar}{4k_B \tau} \tag{20}.$$

Here, $\tau^{-1}$ is the scattering rate, which is given by the $\rho_0$ as

$$\tau^{-1} = -\frac{ne^2 \Delta \rho_0}{m^*} \tag{21},$$

where $m^*$ is the effective mass with the value between $2m_e$ and $4m_e$, and it should be temperature-independent regardless of carriers.

**Figure 21** shows the normalized $T_c$ as a function of differential residual resistance ($\Delta \rho_0$) for $BaFe_{1.89-2x}Zn_{2x}Co_{0.11}As_2$ ($x$ = 0-0.08) [50], $Ba_{0.5}K_{0.5}Fe_{2-2x}Zn_{2x}As_2$ ($x$=0-0.15) [49], $K_{0.8}Fe_{2-y-x}Zn_xSe_2$ ($x$ = 0 and 0.005) [58], $BaFe_{1.85}Co_{0.15}As_2$ under proton-particle irradiation [84], and $NdFeAsO_{0.7}F_{0.3}$ under different $\alpha$-particle irradiation [82], $BaFeAs_{2-x}P_x$ and $BaFe_{1.76}Ru_{0.24}As_2$ under electron irradiation [87,88]. For the chemical doping case like $BaFe_{1.89-2x}Zn_{2x}Co_{0.11}As_2$ calculated from the data in **Figure 17**, the $\Delta T_c$ vs. $\Delta \rho_0$ is observed as a nearly linear relation. Similar relation was also found in the hole-doped $Ba_{0.5}K_{0.5}Fe_{2-2x}M_{2x}As_2$ ($M$ = Mn, Co, Ni, Cu and Zn) system [50]. Considering the linear condition of **Eq. (17)**, the chemical doping results may suggest a weakly change of effective carrier density from doping. For the $\alpha$-particle irradiated $NdFeAsO_{0.7}F_{0.3}$, the $T_c$ slopes nonlinearly with increasing of $\Delta \rho_0$, owing to that the irradiation shifted the chemical potential and enhanced the effective carrier density in linear with irradiation dose, which was demonstrated by Hall effect measurements [82]. We emphasize that the slopes of $T_c$-$\Delta \rho$ are mostly located within the gray region for $BaFe_{1.89-2x}Zn_{2x}Co_{0.11}As_2$, $Ba_{0.5}K_{0.5}Fe_{2-2x}Zn_{2x}As_2$, the $\alpha$-particle irradiated $NdFeAsO_{0.7}F_{0.3}$, and electron irradiated $Ba_{1-x}K_xFe_2As_2$. The slope of $K_{0.8}Fe_{2-y-x}Zn_xSe_2$ may also locate on the gray region, because the Zn doping level of $x$ = 0.005 cannot be considered as the critical point, which is probably pronouncedly less than the experimental result. Note that under the irradiation of light particles like protons and electrons, various 122-type superconductors exhibit relatively larger slopes than other crystals, including the proton-irradiated $n$-type $BaFe_{1.85}Co_{0.15}As_2$, and the electron-irradiated $p$-type $Ba_{1-x}K_xFe_2As_2$, isovalence-type $BaFeAs_{2-x}P_x$, and $BaFe_{1.76}Ru_{0.24}As_2$. However, it is a big challenge to access the critical $\Delta \rho$ for an exact evaluation of the slop, due to the weak power of the irradiation. We emphasize that the recent measurements on the Ba$_{0.5}$K$_{0.5}$Fe$_{2-2x}$Zn$_{2x}$As$_2$ microbridges (BaK-Zn-2) demonstrate much smaller slope of $T_c$-$\Delta\rho$ than the bulk crystals (BaK-Zn), although using the same materials. The pronounced reduction of $\Delta\rho$ can be attributed to the improved characteristic method which can provide accurate evaluation for the pairing-breaking effect (see **Figure 19**).

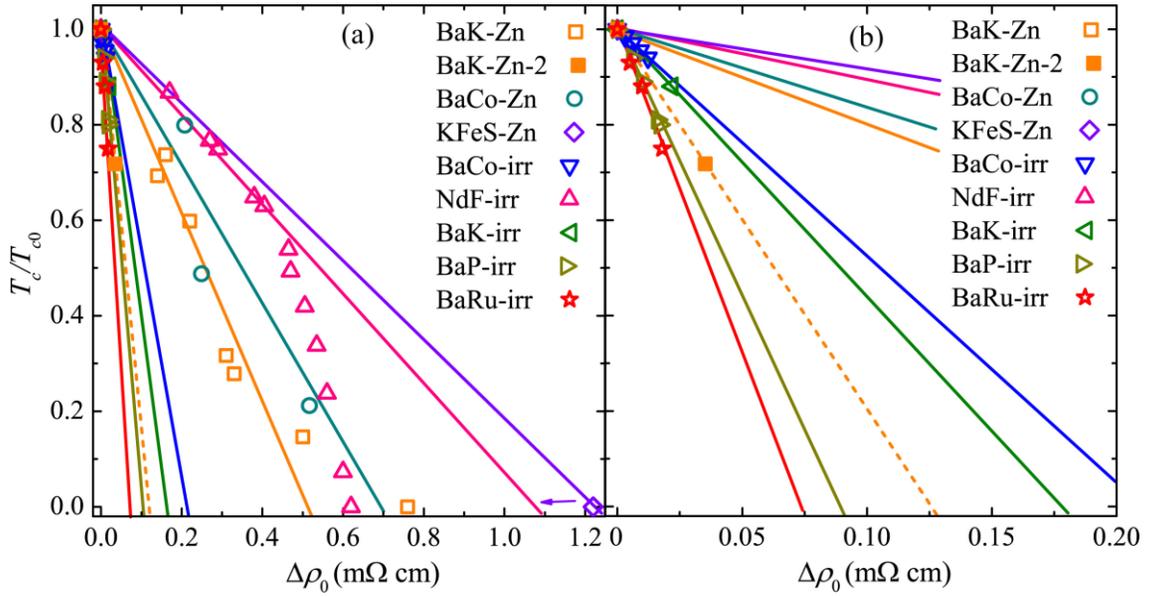

**Figure 21**. (a) Normalized $T_c$ as a function of differential residual resistance ($\Delta\rho_0$) for single crystals BaK-Zn (Ba$_{0.5}$K$_{0.5}$Fe$_{2-2x}$Zn$_{2x}$As$_2$, $x$=0-0.15) [50], BaCo-Zn (BaFe$_{1.89-2x}$Zn$_{2x}$Co$_{0.11}$As$_2$, $x$ = 0-0.08) [49], KFeS-Zn (K$_{0.8}$Fe$_{2-y-x}$Zn$_x$Se$_2$, $x$ = 0 and 0.005) [58], BaCo-irr (BaFe$_{1.85}$Co$_{0.15}$As$_2$ under different proton-particle irradiation) [84], NdF-irr (NdFeAsO$_{0.7}$F$_{0.3}$ under different $\alpha$-particle irradiation) [82], BaK-irr (Ba$_{1-x}$K$_x$Fe$_2$As$_2$ under electron irradiation) [86], BaP-irr (BaFeAs$_{2-x}$P$_x$ under different electron irradiation) [87], and BaRu-irr (BaFe$_{1.76}$Ru$_{0.24}$As$_2$ under different electron irradiation) [88]. BaK-Zn-2 (Ba$_{0.5}$K$_{0.5}$Fe$_{2-2x}$Zn$_{2x}$As$_2$, $x$=0 and 0.05) is measured from microbridges as shown in **Figure 20** [90]. All data are fitted linearly and marked in the same color as the data. The grey region demonstrates the located region for most points. (b) The enlarged view of the low $\Delta\rho_0$ region.

### *(iv) Residual resistance per plane*

The residual resistance may vary from crystals, and some intrinsic defects as grain boundary or magnetic domain may contribute additional scattering as well, thus one can hardly obtain accurate determinations of scattering rate directly from the resistivity results. Instead, further seeking the residual resistance per plane ($\Delta\rho_\square$) is a promising approach. In a 2D system, the $\Delta\rho_\square$ can provide crucial information for the scattering rate of conduction electrons by the impurity potential. $\Delta\rho_\square$ can be described as $\Delta\rho_\square = \Delta\rho_0/t$, where $t$ corresponds to the distance between the nearest neighboring SC layers. Once $\Delta\rho_\square$ reaches to the limit of metallic conductivity $R_q = h/4e^2 = 6.45$ k$\Omega$, the metallic nature of the material will be depressed, as a result of the electron localization. Such method was successfully applied to analyze the cuprate system, for instance, YBa$_2$Cu$_6$O$_{7-\delta}$ [93] and Bi$_2$Sr$_2$Y$_x$Ca$_{1-x}$Cu$_2$O$_8$ [20], whose superconductivity was completely suppressed when $\Delta\rho_\square$ exceeds 6.8 and 8.0 k$\Omega$, respectively.

**Figure 22** gives the normalized $T_c$ as a function of $\Delta\rho_\square$ for BaFe$_{1.89-2x}$Zn$_{2x}$Co$_{0.11}$As$_2$ ($x$ = 0-0.08) [50], Ba$_{0.5}$K$_{0.5}$Fe$_{2-2x}$Zn$_{2x}$As$_2$ ($x$=0-0.15) [49], K$_{0.8}$Fe$_{2-y-x}$Zn$_x$Se$_2$ ($x$ = 0 and 0.005) [58], BaFe$_{1.85}$Co$_{0.15}$As$_2$ under different proton-particle irradiation [84], NdFeAsO$_{0.7}$F$_{0.3}$ under different $\alpha$-particle irradiation [82], Ba$_{1-x}$K$_x$Fe$_2$As$_2$ ($x$=0.19, 0.26, and 0.34), BaFeAs$_{2-x}$P$_x$, and BaFe$_{1.76}$Ru$_{0.24}$As$_2$ under electron irradiation [87,88,90]. For the 122-system, regardless of $p$-type or $n$-type, substitution of Zn enhances the $\Delta\rho_\square$ remarkably and depresses the superconductivity at 6 k$\Omega$. The $\alpha$-particle irradiated NdFeAsO$_{0.7}$F$_{0.3}$ single crystal demonstrates a slightly larger critical point of $\Delta\rho_\square$ (~7.5 k$\Omega$). In addition, the K$_{0.8}$Fe$_{2-y-x}$Zn$_x$Se$_2$ exhibits intense increase of $\Delta\rho_\square$ (~8.7 k$\Omega$) with only 0.25at.% of Zn-doping, where the superconductivity was completely depressed. Indeed, the critical point of $\Delta\rho_\square$ may be less than the present values, due to lack of intermediate doping levels. The superconductivity of $\alpha$-particle irradiated NdFeAsO$_{0.7}$F$_{0.3}$ and K$_{0.8}$Fe$_{2-y-x}$Zn$_x$Se$_2$ is suppressed at the metallic limit point 6.45 k$\Omega$, suggesting the possible effect from electron localization. In contrast, the proton irradiated BaFe$_{1.85}$Co$_{0.15}$As$_2$ single crystal, electron-irradiated Ba$_{1-x}$K$_x$Fe$_2$As$_2$, BaFeAs$_{2-x}$P$_x$, and BaFe$_{1.76}$Ru$_{0.24}$As$_2$ single crystals, and Ba$_{0.5}$K$_{0.5}$Fe$_{2-2x}$Zn$_{2x}$As$_2$ microbridges exhibit considerably smaller critical points below 1.5 k$\Omega$, being far away from 6.45 kΩ. However, another important parameter should be taken into account, i.e., the carrier density ($n$). For instance, in the electron-type superconductors, the electron-provider, e.g. Co, contributes to the scattering as well. Therefore, the scattering factor can be revised as $\Delta\rho_\square n$, for which the Hall coefficient measurements are essential.

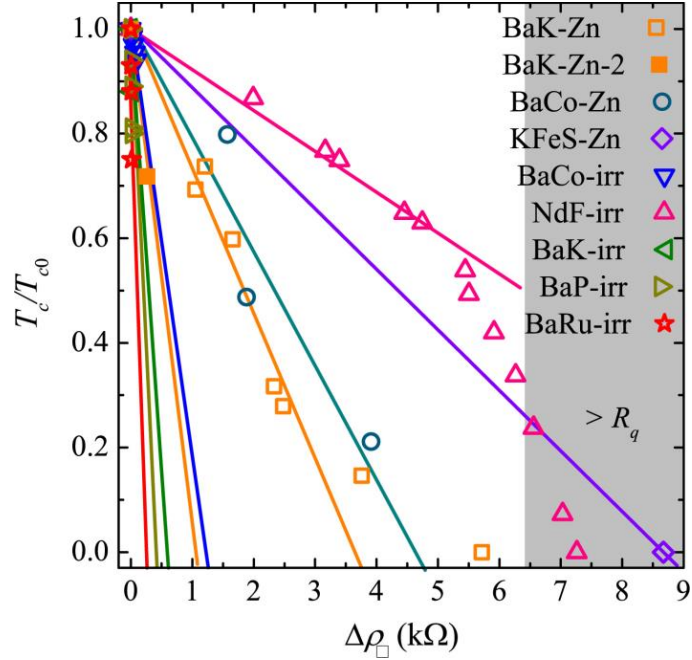

**Figure 22**. Normalized $T_c$ as a function of residual resistance per plane ($\Delta\rho_\square$) single crystals BaK-Zn ($Ba_{0.5}K_{0.5}Fe_{2-2x}Zn_{2x}As_2$, $x$=0-0.15) [50], BaK-Zn-2 ($Ba_{0.5}K_{0.5}Fe_{2-2x}Zn_{2x}As_2$, $x$=0 and 0.05) microbridges [90], BaCo-Zn ($BaFe_{1.89-2x}Zn_{2x}Co_{0.11}As_2$, $x$ = 0-0.08) [49], KFeS-Zn ($K_{0.8}Fe_{2-y-x}Zn_xSe_2$, $x$ = 0 and 0.005) [58], BaCo-irr ($BaFe_{1.85}Co_{0.15}As_2$ under different dose of proton-particle irradiation) [84], NdF-irr ($NdFeAsO_{0.7}F_{0.3}$ under different dose of $\alpha$-particle irradiation) [82], BaK-irr ($Ba_{1-x}K_xFe_2As_2$ under different dose of electron irradiation) [86], BaP-irr ($BaFeAs_{2-x}P_x$ under different dose of electron irradiation) [87], and BaRu-irr ($BaFe_{1.76}Ru_{0.24}As_2$ under different dose of electron irradiation) [88]. All data are fitted in linear and marked in the same color as the data. The grey region demonstrates resistance larger than the $R_q$=6.45 kΩ.

*(v) Carrier density*

Different from the normal metal, the Hall coefficient of FBS displays a temperature dependent behavior, which is one of the important signatures of the normal state as those of the cuprate systems. In the cuprate superconductors, Anderson [94-96] proposed a unconventional approach to deal with the anomalous Hall effect, namely, the NMI doping effect on the carrier density and Hall angle. As substitution for Fe atoms in the $Fe_2X_2$ plane, the Zn impurity provides a large negative scattering potential (around -8 eV [10]). However, the Zn substitution negligibly changes the electronic structure, which was supported by the $^{75}$As NMR and NQR measurements on La(Fe$_{1-x}$Zn$_x$)AsO$_{0.85}$ polycrystal [79]. In addition, the $^{139}$La NMR spectra indicated that the substituted Zn could not cause any static moments around the impurity [79], different from that of cuprate systems in which both local spin susceptibility and staggered susceptibility are prominently enhanced around the Zn site, within a radius of the antiferromagnetic correlation length $\xi_{AF}$.

Up to now, apart from few reports on Hall effect measurements on the Zn-doped polycrystals, there are only studies on in-plane Hall effect on the 122-type single crystals, namely, the $n$-type BaFe$_{2-2x-2y}$Zn$_{2x}$Co$_{2y}$As$_2$ and $p$-type Ba$_{0.5}$K$_{0.5}$Fe$_{2-2x}$M$_{2x}$As$_2$ ($M$ = Fe, Mn, Ru, Co, Ni, Cu and Zn) as shown in **Figure 23** [49]. The data of the impurity-free crystals are compatible with the earlier data. Regarding the normal state, no significant change of Hall coefficient is induced by the Zn substitution in the under-, optimal-, and over-doped regimes of Co in the $n$-type BaFe$_{2-2x-2y}$Zn$_{2x}$Co$_{2y}$As$_2$, indicating that the substitution is truly isovalent which negligibly change the carrier density. The weak carrier density change in the $p$-type Ba$_{0.5}$K$_{0.5}$Fe$_{2-2x}$Zn$_{2x}$As$_2$ Was likely caused by transferring of carrier type from hole to electron since it is anomalous that the Zn impurities resulted in a negative $R_H$. Here, it should be noted that the present system has the multi-band nature with both electron and hole carriers. The results for the Zn doped samples may suggest that the electron carrier become dominant with existence of the substitution. However, such Hall experiments were implemented on a bulk crystal, while the Hall resistance was generally rather small owing to the bulk size and in-plane metallic behavior, thus we can hardly understand the nature of Zn impurities effect on the in-plane scattering.

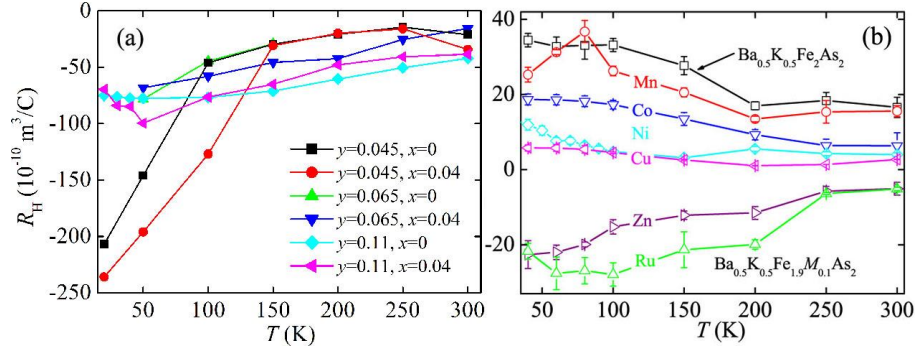

**Figure 23**. Temperature dependence of in-plane Hall coefficient ($R_H$) for single-crystalline (a) $BaFe_{2-2x-2y}Zn_{2x}Co_{2y}As_2$ and (b) $Ba_{0.5}K_{0.5}Fe_{2-2x}M_{2x}As_2$ ($M$ = Mn, Ru, Co, Ni, Cu and Zn, nominal $x$=0.05), from Li e$t$ $al$. [49].

To enhance the Hall signal we recently developed a method to measure the transport properties on single crystalline microbridges as shown in **Figure 19**. Since the thickness of the microbridge can be down to 91 nm or less, thus the in-plane resistance and Hall resistance (under magnetic field of 5 T) at room temperature were up to 20 Ω and 0.5 Ω, respectively, which provides an impressive way to study the accurate transport properties of the metallic Fe-based superconductors. **Figure 24** shows contours of $R_H$ for $Ba_{0.5}K_{0.5}Fe_2As_2$ (BK), $Ba_{0.5}K_{0.5}Fe_{1.95}Zn_{0.05}As_2$ (BKZn), and $Ba_{0.5}K_{0.5}Fe_{1.95}Co_{0.05}As_2$ (BKCo) measured in various magnetic fields and temperatures regions. For the BK, $R_H$ shows two pronounced $H$-dependence regions, those are, an $H$-dependent region above 250 K and an $H$-independent below ~ 250 K as described above. In the $H$-independent region, the $R_H$ behaves positive indicating hole-type carriers, while a negative $R_H(H)$ phenomenon is observed at relatively low field condition in the $H$-dependent region, suggesting a Hall sign reversal occurs which seems like that the electrons dominate the carriers. However, further increasing of field turns the sign back to positive. Meanwhile, the negative $R_H(H)$ appears in a narrow temperature region between 290 and330 K.

As substituting of nonmagnetic Zn ions into the BK, the $R_H$ is increased to about twice of the original magnitude, and the field dependent behavior is modified as well, i.e., the $R_H$ is independent of field in the whole temperatures region and the negative $R_H$ disappears, indicating $H$-independent carriers at all temperatures. For the magnetic impurity Co substituted BK, the $R_H$ becomes more complicated: (i) the $R_H$ demonstrates a pronounced increase; (ii) $R_H$ reveals intense $H$-dependent behavior at high temperature, and the $T_H$ (~ 240 K) is slightly smaller than that of BK; and (iii) the negative $R_H$ regime exists at almost the whole $H$-dependent region, and the saturated high field is essential to suppress the negative $R_H$. These data suggest that the carriers are electron-type at almost all fields above the $T_H$ (240 K). However, the carriers of both BKZn and BKCo are similar with those of BK below the $T_H$, which seems like that the impurities induce modification only on electron-pocket but not on the hole-pocket. To confirm the possible effect of $Co^{2+}$ and $Zn^{2+}$ on the electronic state, it is necessary to study the Hall angle for both impurity-free and impurity-doped samples.

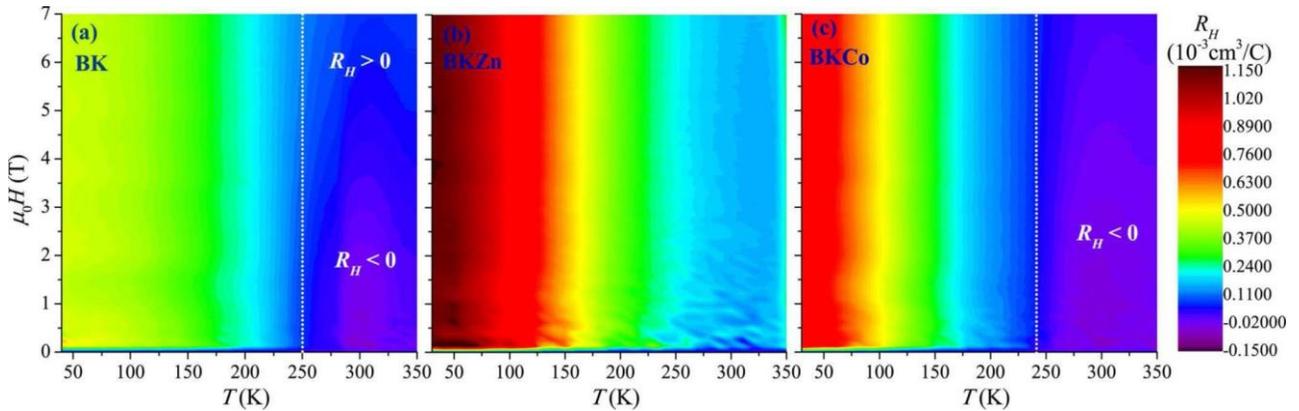

**Figure 24**. Contours of $R_H$ for $Ba_{0.5}K_{0.5}Fe_2As_2$ (BK), $Ba_{0.5}K_{0.5}Fe_{1.95}Zn_{0.05}As_2$ (BKZn), and $Ba_{0.5}K_{0.5}Fe_{1.95}Co_{0.05}As_2$ (BKCo) with respect of magnetic fields and temperatures. The white dot line separates two temperature regions demonstrating linear (left side) and non-linear (right side) field dependent regions for $R_H(H)$, where we define the crossover temperature as $T_H$ (~ 250 K for BK and ~240 K for BKCo), which separates the two temperature regions as $H$-dependent and $H$-independent $R_H(H)$, from Li *et al*. [90].

*(vi) Hall angle*

To explore the impurity effects on the electronic state, Anderson proposed a picture for describing the Hall angle (cot$\theta_H$) within the framework of the Luttinger-liquid theory which can be expressed as [94-96]

$$\cot\theta_H = \frac{\rho_{xx}}{\rho_{xy}} = \alpha T^2 + C \tag{22},$$

where $\rho_{xx}$ and $\rho_{xy}$ are the longitudinal and transversal resistivity, respectively, and $\alpha$ is a parameter dependeding on the energy-scale of the spinon-spinon scattering

$$\alpha = \frac{m_s}{eHW_s} \tag{23},$$

where $m_s$ and $W_s$ are effective mass and the bandwidth of spin excitations, respectively, and $C$ is the in-plane impurity contributed scattering rate

$$C = \frac{m_s}{eH\tau_M} \tag{24}$$

in which $\tau_M$ is the impurity contribution. Therefore, the cot$\theta_H$ depends only on the transverse scattering lifetime ($\tau_H$), which is determined mainly by spinon-spinon interactions and characterized by a $T^2$ dependence. Among the appealing features of the theory few parameters are required to describe the normal-state transport, and the understanding of the general features of the normal-state properties does not involve details of the electronic structure. In the cuprate superconductors, for instance YBCO, the quantity $C$ behaves a linear function of Zn-doping level ($x$) while $\alpha$ is a constant [97, 98], the slops of various crystals are consequently a constant regardless of the Zn-doping levels.

**Figure 25(a)** shows the $T^2$ dependence of cot$\theta_H$ for Ba$_{0.5}$K$_{0.5}$Fe$_2$As$_2$ (BK), Ba$_{0.5}$K$_{0.5}$Fe$_{1.95}$Zn$_{0.05}$As$_2$ (BKZn), and Ba$_{0.5}$K$_{0.5}$Fe$_{1.95}$Co$_{0.05}$As$_2$ (BKCo) superconductors, which are abbreviated as BK, BKZn and BKCo, respectively. All crystals exhibit a non-linear change with $T^2$ at the high-$T$ region. For instance, for BK the non-linear $T^2$ dependence appears above 250 K, suggesting a larger power of $T^n$ with $n$>2. The crossover temperature, however, is well accordance with both $T^*$ and $T_H$ as discussed above. As nonmagnetic Zn substitution, the cot$\theta_H$ shows weak non-linear change at high-$T$. For the magnetic Co impurity the cot$\theta_H(T^2)$ however indicates a dramatically modification from that of BK at high-$T$. Since the Hall signals are rather small at this temperature region, where sign reversal takes place as discussed above, we can hardly acquire reliable information from these results. We then focus on the low-$T$ regions where the cot$\theta_H$ demonstrates linear change with $T^2$. **Figure 25(b)** gives the linear fitting of $T^2$ for cot$\theta_H$ of BK, BKZn and BKCo at $T < 140$ K. The $\alpha$ ($C$) is observed as $2.69 \times 10^{-3}$ (3.96), $2.43 \times 10^{-3}$ (15.38) and $3.48 \times 10^{-3}$ (12.84) for BK, BKZn and BKCo, respectively. The Zn seems unlikely to change the slope of cot$\theta_H$ from that of BK, which could be understood in terms of Eq.(22). The behavior of cot$\theta_H$ is consistent with the intrinsic relaxation rate of the elementary excitations (spinons) in the normal state, but not a "transverse" relaxation rate. Indeed, the weak modification of Zn of the spin state was confirmed by the $^{75}$As NMR and NQR measurements on La(Fe$_{1-x}$Zn$_x$)AsO$_{0.85}$ polycrystal [2]. The $^{139}$La NMR spectra as well indicated that the substituted Zn could hardly induce static moments around the impurity [2]. On the other hand, the Zn substitution increases the in-plane impurity scattering rate dominated $C$ (15.38) of BK (3.96). It is reasonably understood because the Zn provides a large negative scattering potential (around -8 eV [1]) within the Fe$_2$As$_2$ plane. The Zn increases the residual resistivity in linearity and consequently suppresses the superconductivity as described in **Figure 21** and **22**. Substitution of Zn on YBCO demonstrated a similar parallel shift of the cot$\theta_H$ vs.$T^2$ curves with different doping levels [97, 98].

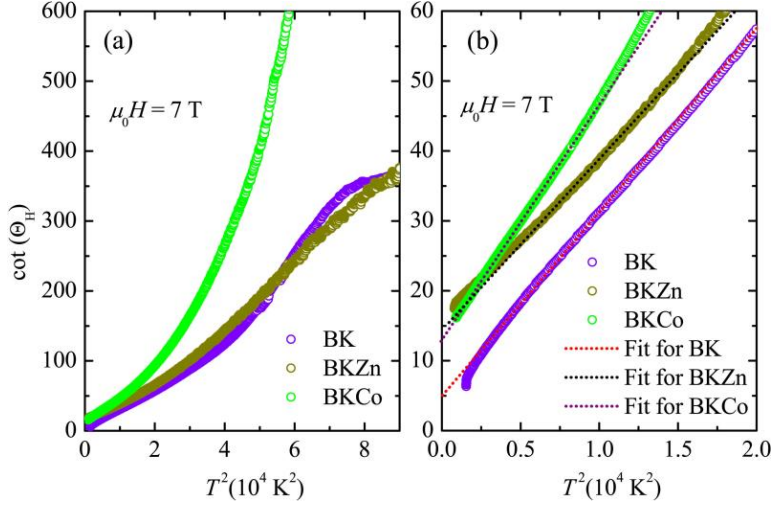

**Figure 25.** The Hall angle $\cot\theta_H = \rho_{xx}/\rho_{xy}$ for a systematic set of BK, BKZn and BKCo. The high temperature data can be easily fitted by a square law $\cot\theta_H = \alpha T^2 + C$, from Li *et al.* [90].

## 4.5. Heat capacity

It is essential to understand whether a NMI can induce a local moment on its vicinity, especially in the low enough temperature regions. The transport properties measurements can hardly be carried out below $T_c$, unless using extremely high fields, large current density, or heavy level impurity doping. Technically, the upper critical fields $H_{c2}$ for FBS are quite high, for instance, the $H_{c2}$ for $Ba_{0.5}K_{0.5}Fe_2As_2$ was found as around 55 T [99]. In other case, the heavy level impurity doped crystals often confront with disorder and inhomogeneous distribution problem. STM measurement is the best method to explore the low-$T$ local moment of Zn as discussed in detail in **Section** 3. Alternatively, low-$T$ heat capacity measurement is a simple approach owing to the bulk properties. From low-$T$ heat capacity measurements, applying a magnetic field will split the free Zn-induced moments into a two-level system and a Schottky anomaly should appear [100,101].

A recent study on optimal $Fe_{1+y}Te_{1-x}Se_x$ ($x \approx 0.4$) showed that the Co or Cu substitution for Fe mainly serves as scatters rather than charge carrier doping [102]. In comparison with Cu doping, the Co substitution shows a less evident suppression effect on superconductivity while it exhibits a stronger influence on magnetism. Upon substitution of Co for Fe, a Schottky heat capacity anomaly develops at low temperatures, implying the existence of a paramagnetic moment, as displayed in **Figure 26**. In contrast, Cu substitution may mainly serve as non-magnetic scatters, where no Schottky anomaly is observed. The results qualitatively support the $s_{\pm}$ wave symmetry of the order parameter.

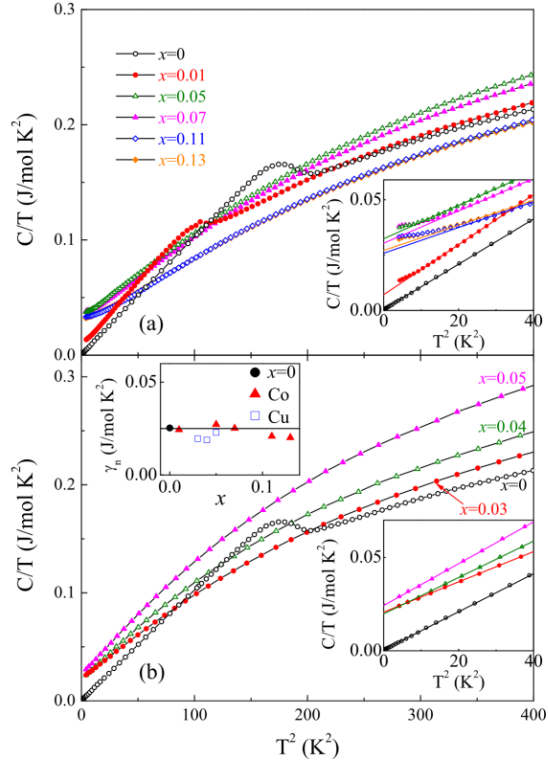

**Figure 26**. The C/$T$ versus $T^2$ plots (a) for Co$_x$Fe$_{1+y-x}$Te$_{0.6}$Se$_{0.4}$ and (b) for Cu$_x$Fe$_{1+y-x}$Te$_{0.6}$Se$_{0.4}$ from Zhang *et al.* [102].

## 4.6. Penetration depth

In addition, the behavior of SC suppression may be determined by a wide range of other possibilities, for instance another low-temperature property like the London penetration depth $\lambda$, when disorder is systematically increased. To make more precise statements, Wang *et al.* [42] proposed some independent ways to fix the scattering potential of a given impurity, and in particular the relative proportion of inter- to intra-band scattering.

Cho and co-workers [86] studied the $\lambda(T)$ of electron-doped $Ba_{1-x}K_xFe_2As_2$ single crystals. The low-temperature behavior of $\Delta\lambda(T)$ was found as a power-law function

$$\Delta\lambda(T) = A\left(\frac{T}{T_c}\right)^n \tag{25},$$

where $A$ is a constant parameter depending on the compound. **Figure 27** shows fitting results for the $\Delta\lambda(T)$ of the $e$-irradiated $Ba_{1-x}K_xFe_2As_2$ single crystals. To eliminate the uncertainty related to the upper fitting limit, several fittings were applied with a variable high-temperature end of the fitting range, namely, $T_{up}/T_c \sim 0.1$ to $0.3$, while keeping the lower limit at the base temperature. The heavily doped samples exhibit strong saturation behavior with the large exponent values, $n > 3$. For the samples close to the optimal doping, namely, $x = 0.24$ and $0.32$, the exponent $n$ remained high ($n = 3$) and increases with decreasing $T_{up}/T_c$. This implies that the temperature dependent $\lambda(T)$ remains exponential at low temperatures with the addition of nonmagnetic scattering, and consequently supports the isotropic gap like single-band $s$-wave or multiband $s_{++}$ wave. For the most under-doped state with $x = 0.19$, there is a clear evolution toward the $T^2$ relation, which reveals the changes from exponential in the clean limit to $\sim T^2$ in the dirty limit, and supporting the nodeless $s_{\pm}$ pairing.

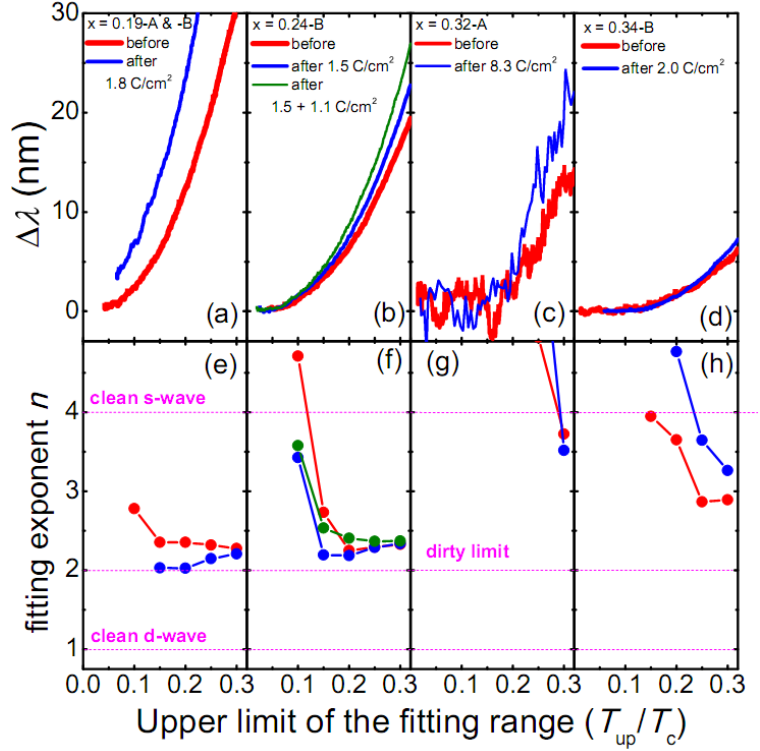

**Figure 27**. Temperature dependence of $\Delta\lambda$ at the low-temperature region for Ba$_{1-x}$K$_x$Fe$_2$As$_2$ single crystals with and without electron irradiation. The data were fitted with a power-law function as Eq. (25). (e)-(h) Corresponding to fit exponents with fitting errors less than $\pm 0.1$, from Cho *et al*. [86].

## 4.7. Electronic thermoelectric power

The transport properties, for instance the Hall effect, is measured for the whole compound, but some compounds, like BaFe$_{2-2x-2y}$Zn$_{2x}$Co$_{2y}$As$_2$, have already doped with other elements that will strongly influence the Hall effect. Therefore, two scattering rates occur unless for an "ideal" pure single crystal. However, it is even harder to perform quantitative comparisons between samples than for the residual resistivity. Thus, the data can hardly provide the determination of the physical origin for the actual difference between the two scattering rates alone, for which some additional experiments such as electronic thermoelectric power on the Zn-substituted single crystals will be greatly helpful. In this case we can systematically study the electronic thermoelectric power in SC states of various Zn-doped compounds, and explore if Zn has contributed to the normal-state electronic or SC states, as what were comprehensively studied in cuprates [103].

# 5. Impurities in superconducting state

## 5.1. 11-system

The 11-system is featured by without the separating blocking-layer among the FBS family. The substitution of impurity is indeed onto the Fe-site rather than other sites within the blocking-layer as that occurs in other systems. Zhang and co-workers showed that among the $3d$ transition metals from Cr to Zn, only the Co, Ni, and Cu with smaller ionic radii for valence state $2^+$ can substitute effectively for Fe in $Fe_{1+y}Te_{1-x}Se_x$ ($x \approx 0.2$) [102]. The substitutions of any $3d$ metals with the concentration of ~5% can lead to the formation of spin-glass state and suppression of superconductivity. As shown in **Figure 28**, in contrast to the Co substitution which slightly suppresses the superconductivity, the Cu and Ni substitutions with the same concentration completely destroy the superconductivity.

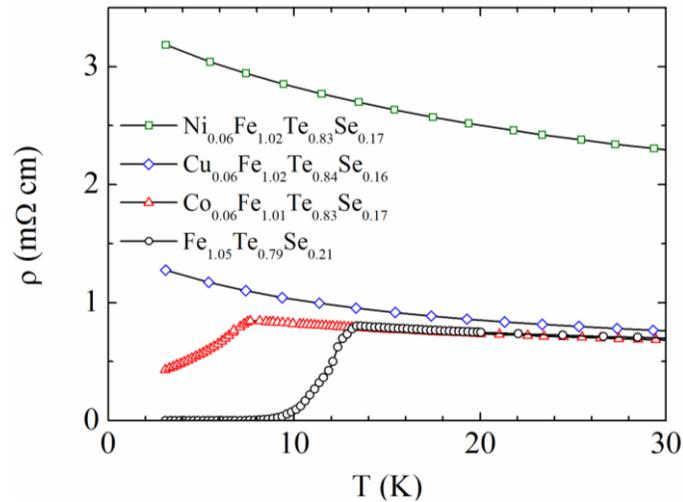

**Figure 28**. In-plane resistivity as functions of temperature for $M_z Fe_{1+y-z}Te_{1-x}Se_x$ ($M$=Co, Ni and Cu, $z = 0, 0.06, x \approx 0.2$), from Zhang *et al*. [102].

## 5.2. 1111-system

As briefly introduced in **Sec 1**, the Fe-based 1111- and 122-system are two preferred systems for study on the NMI substitution effects. The Zn-doping study was firstly carried out on the 1111-system which possesses the highest $T_c$ among the FBS. The first observation of $T_c$ suppression was reported by Guo *et al*. in polycrystalline Zn-doped LaFeAsO$_{0.85}$ [48]. **Figure 29** presents the magnetic susceptibilities of the LaFe$_{1-x}$Zn$_x$AsO$_{0.85}$, which manifest a large $T_c$ decrease from 26 K to zero only by a minimal amount of Zn (< 3 at. %) substitution. The results likely indicate that the conventional *s*-wave model is highly unlike for LaFeAsO$_{0.85}$, but the $s_\pm$ wave model and the nodal *d*-wave model both remain possible. Li and co-workers [104, 105] also studied the Zn substitution effects on SC of the polycrystalline LaFe$_{1-y}$Zn$_y$AsO$_{1-x}$F$_x$. As is shown in **Figure 30**, the $T_c$ suppression by Zn was only observed in the fluorine over-doped regime, whereas in the fluorine under- and optimal-doped regimes, the $T_c$ was even slightly enhanced which was unusual and had never been found in any other superconductors. Seen from the temperature dependent resistivity of the fluorine under-doped samples, the low-$T$ upturn was gradually suppressed by Zn impurities and the carrier density was also enhanced indicated by the Hall measurements. Apart from the influence of polycrystalline nature of samples, one can hardly explain the Zn effects according to the discussion in **Section 4.4**. It seems like that the Zn substitution might improve the disorder or other intrinsic reasons. Later theoretical study from their colleagues suggested that the presence of a Zn impurity could induce an electron transferring from As to Fe atoms in both the fluorine under-doped and over-doped regions and modify the local lattice structure. Further experiments on the single crystals could be helpful to elucidate this phenomenon.

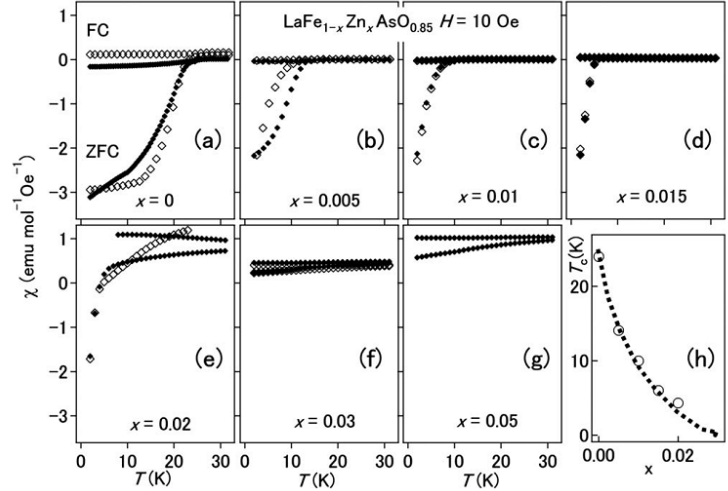

**Figure 29.** $T$ and $x$ dependence of $\chi$ of LaFe$_{1-x}$Zn$_x$AsO$_{0.85}$ measured at 10 Oe under ZFC and FC conditions. Open and closed symboles represent data from independent set of samples. (h) $T_c$ vs. $x$, from Guo *et al.* [48].

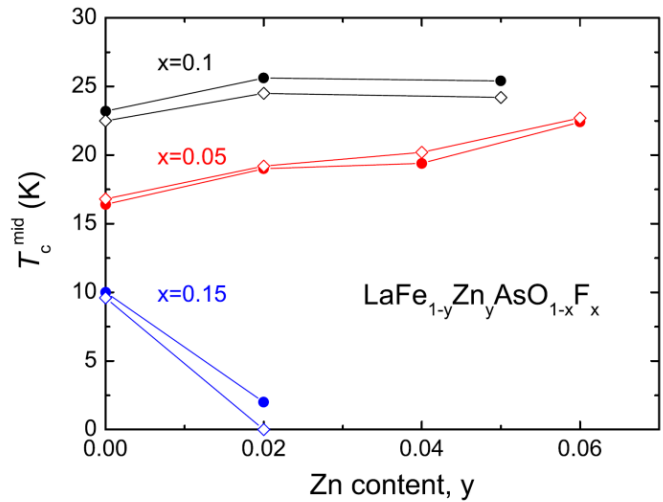

**Figure 30.** $T_c$ as the function of Zn content in polycrystalline LaFe$_{1-y}$Zn$_y$AsO$_{1-x}$F$_x$. Solid and open symbols refer to $T_c$ determined from resistivity (midpoint) and susceptibility (onset point) measurements, respectively, from Li e*t al.* [104].

Li and co-workers also studied the Zn-substitution effect on the *n*-type 1111-system [106]. **Figure 31** shows the Zn content dependent $T_c$ of polycrystalline LaFe$_{1-x-y}$Co$_y$Zn$_x$AsO in the cobalt

under-, optimal- and over-doped regimes. Different from their previous results on the *p*-type LaFe$_{1-y}$Zn$_y$AsO$_{1-x}$F$_x$, SC in these samples was suppressed by Zn doping in all cobalt-doped regimes. The SC disappears at Zn content of about 12at.%, 4at.%, and 3at% for the cobalt under-, optimal- and over-doped regimes, respectively. The slopes of $T_c(x)$ are similar for the optimal- and over-doped regimes and both of them are more sharply than that of under-doped case.

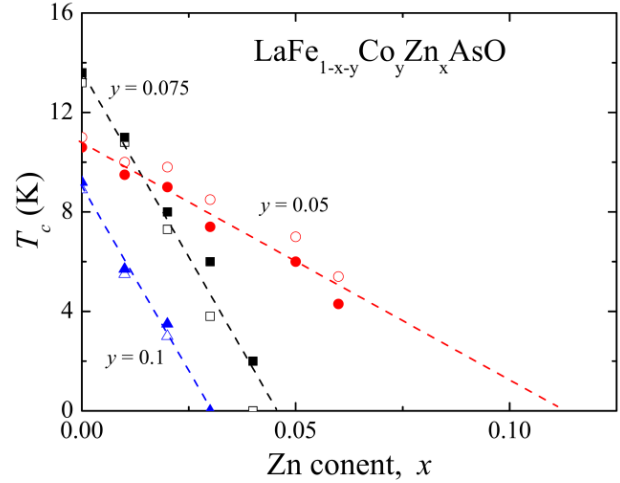

**Figure 31.** Zn content dependence of $T_c$ of polycrystalline LaFe$_{1-x-y}$Co$_y$Zn$_x$AsO. Solid and open symbols refer to $T_c$ determined from resistivity (midpoint) and susceptibility (onset point) measurements, respectively, from Li e*t al*. [106].

In addition, Sato and co-workers suggested that the Co impurity should act as a NMI comparing with the host atom Fe. They studied the substitution effects of Mn and Co on $R$FeAsO$_{0.89}$F$_{0.11}$ ($R$= La and Nd) polycrystalline samples [13]. **Figure 32** gives $T_c$ as a function of the impurity content for $R$Fe$_{1-y}$M$_y$AsO$_{0.89}$F$_{0.11}$ ($R$= La and Nd; $M$ = Co and Mn). In these studies, the Co ion was believed to provide two electrons to the host conduction bands and was characterized as a NMI. Ni [107] and Ru [108] were considered as nonmagnetic as well. The SC was completely suppressed with Co impurity content of 7.5at.%, suggesting that the $T_c$ suppression rate is well accordance with LaFe$_{1-x-y}$Co$_y$Zn$_x$AsO case as we discussed above. It is worth noting that magnetic impurity of Mn introduced more dramatic $T_c$ suppression than any other impurities.

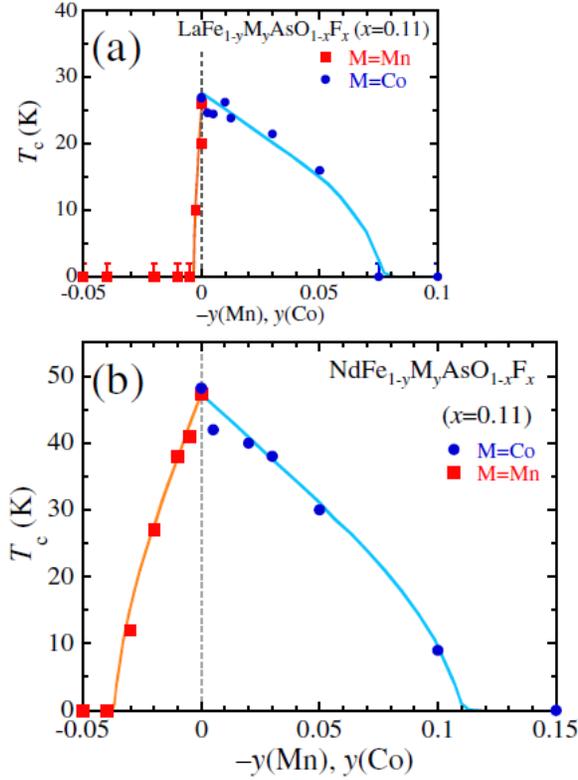

**Figure 32.** The $T_c$ as a function of impurity content for $R$Fe$_{1-y}M_y$AsO$_{0.89}$F$_{0.11}$ polycrystal, where the $R$= La and Nd, and $M$ = Mn and Co, from Sato *et al*. [13].

## 5.3. n-type 122-system

Up to now, most of NMI study on FBS was reported on polycrystals as introduced above. Nevertheless, a single crystal has sizable and atomically ordered surfaces, which is crucial for surface sensitive measurements and scattering behavior within the SC plane, as discussed in **Section 4.1** to **4.3**. Consequently, the growth of Zn-doped single crystals is of great importance to study the intrinsic Zn impurity effects. Since it is rather difficult in the growth of single crystals for the 1111-system, the well-studied 122-system is a promising candidate for the study.

Utilizing the high-pressure technique, Zn impurities were successfully doped into the single crystal in 122-system, as discussed in **Section 4.1**. **Figure 33** summarizes for the results of $T_c$ suppression as a function of the impurity concentration for $BaFe_{2-2x-2y}M_{2x}Co_{2y}As_2$ ($M$ = Zn, Mn) [46]. Here, the $T_c$ is derived from the resistivity. SC is fully suppressed by Zn at the common doping level of ~8 at.%, regardless of the under- ($y = 0.045$), optimal- ($y = 0.055$) and over-doped ($y = 0.11$) regimes. The suppression rate for the Zn doping was roughly estimated by applying a linear function fit to the data, giving 2.31, 3.63, and 2.45 K/at.% for the under-, optimal- and over-doped states, respectively. The $T_c$ decreasing rates by Zn change little from the under- to over-doped SC states, being a remarkable feature discovered in this study. Clearly, the rates are much smaller than the expected one for the $s_{\pm}$ wave model (25 K/at.%, see below), indicating that the observed SC is much robust against the NMI than the expectation for the $s_{\pm}$ wave SC state. Similar results were also found in the Zn-doped $BaFe_{1.92-2x}Zn_{2x}Pt_{0.08}As_2$ single crystals [109]. For a comparison, the data for polycrystalline $BaFe_{1.87-2x}Zn_{2x}Co_{0.13}As_2$ prepared under ambient-pressure are co-plotted by a dotted line in **Figure 32**. The discrepancy between the present and the earlier data is probably due to overestimation of the Zn content in the ambient-pressure sample. Since the Zn substitution is not easy caused by the low melting point and high volatility of Zn-containing compounds, it is uncertain that whether Zn can be substituted into the Fe-site by the normal ambient-pressure synthesis. The suppression rate for the magnetic Mn doped sample was estimated to be 6.32 K/at.%, which is still quite smaller than the theoretical value for the $s_{\pm}$ wave model [10, 11]. **Figure 34** shows the SC phase diagram of $BaFe_{2-2x-2y}Zn_{2x}Co_{2y}As_2$, where $T_c$ is determined by the resistivity measurements. As aforementioned, SC is suppressed by the common doping level of Zn, ~8at.%.

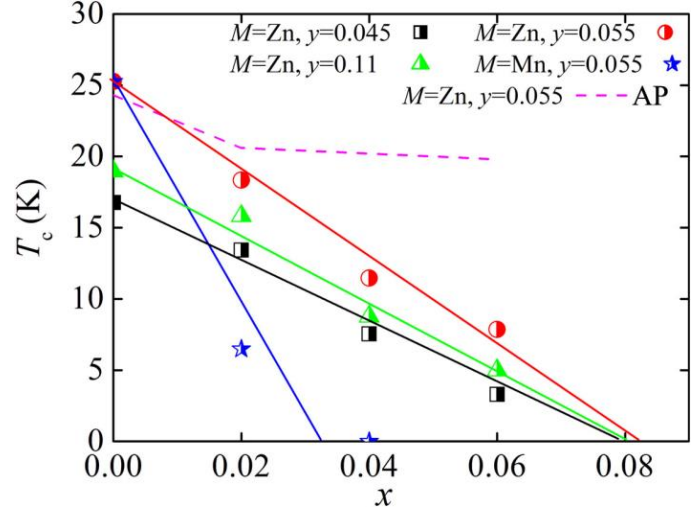

**Figure 33.** Impurity concentration $x$ dependence of $T_c$ for BaFe$_{2-2x-2y}M_{2x}$Co$_{2y}$As$_2$ single crystals, where the impurities $M$ are Mn and Zn, and the $T_c$ were derived from resistivity. The data shown by the broken line are from the samples prepared under ambient-pressure, from Li e*t al*. [46].

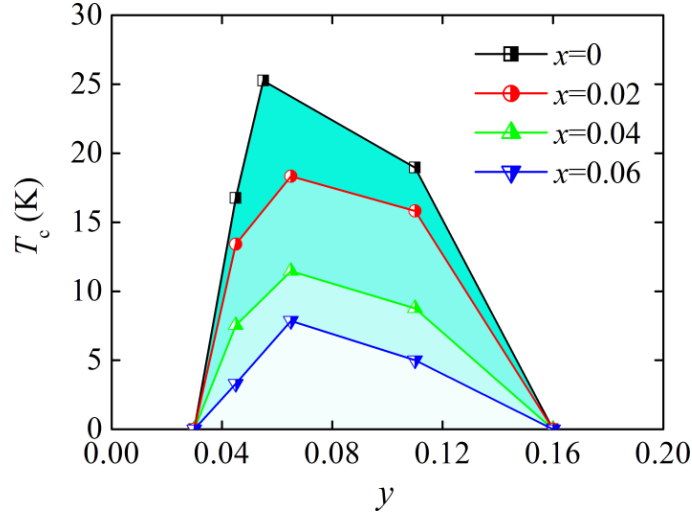

**Figure 34**. Electronic phase diagram for BaFe$_{2-2x-2y}$Zn$_{2x}$Co$_{2y}$As$_2$ single crystals, from Li e*t al*. [46].

Similar $T_c$-suppression behavior was reported by Nakajima *et al*. [84], which was caused by the application of 3 MeV proton irradiation on BaFe$_{2-2x}$Co$_{2x}$As$_2$ single crystals at under-, optimal-, and over-doping levels. **Figure 35** shows the normalized $T_c$ for crystals under different irradiation dose, where $T_{c0}$ is the $T_c$ irradiation-free sample. In all the doping regimes, $T_c/T_{c0}$ decreases linearly with increasing the dose, whereas the slopes are different with respect to the doping regimes, from which the over-doped case was estimated as the most drastic SC suppression effect. In contrast, the $T_c$ in the optimal-doping regime showed relatively robust against irradiated defects than other regimes. Such sample dependent $T_c$ suppression behavior seems like inconsistence with those of chemical doping results as discussed above. Detailed analysis for the $\rho_0$ and pair-breaking will be discussed in **Section** 5.

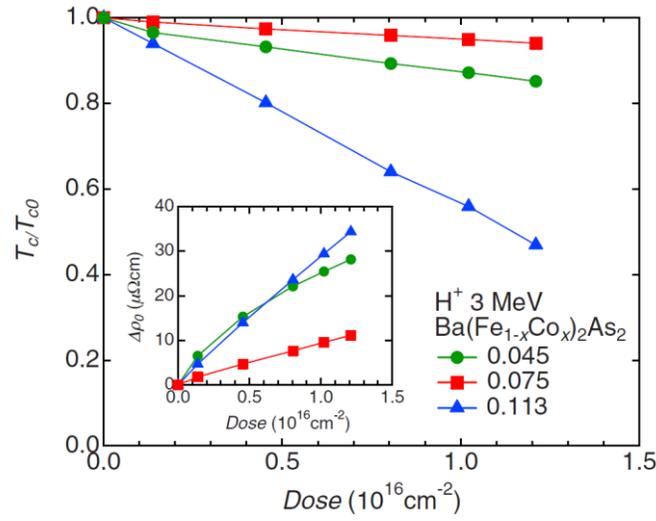

**Figure 35**. Irradiation dose dependence of the normalized critical temperature $T_c/T_{c0}$ for BaFe$_{2-2x}$Co$_{2x}$As$_2$ single crystals. The doping state of Co were in under- ($x = 0.045$), optimal- ($x = 0.075$), and over-doping ($x = 0.113$) regions, whose $T_{c0}$ is 15.1 K, 24.8 K, and 12.8 K, respectively. Inset: $\Delta\rho_0$ as a function of dose, where $\Delta\rho_0 = \rho_0^{irr} - \rho_0^{unirr}$, from Nakajima *et al*. [84].

## 5.4. p-type 122-system

Cheng *et al*. studied the Zn and Mn substitution effects on SC of the *p*-type polycrystalline Ba$_{0.5}$K$_{0.5}$Fe$_2$As$_2$ synthesized under ambient-pressure (see **Figure 36**) [51]. They found that the Mn doping depressed the SC transition temperature drastically with a rate of $\Delta T_c$/Mn=−4.2 K/at.%. Transport properties measurement revealed that the Mn doping enhanced the $\rho_0$ significantly, and induced strong local magnetic moments (~2.58 $\mu_B$) for distinct pair-breaking. Such results were well consistent with those of Mn-doping in other systems [46]. In contrast, the $T_c$ and $\rho_0$ were observed as almost independent of Zn-doping. The weak Zn substitution effect is only similar as that of Zn under- and optimal-doped LaFeAsO$_{1-y}$F$_y$ system [104].

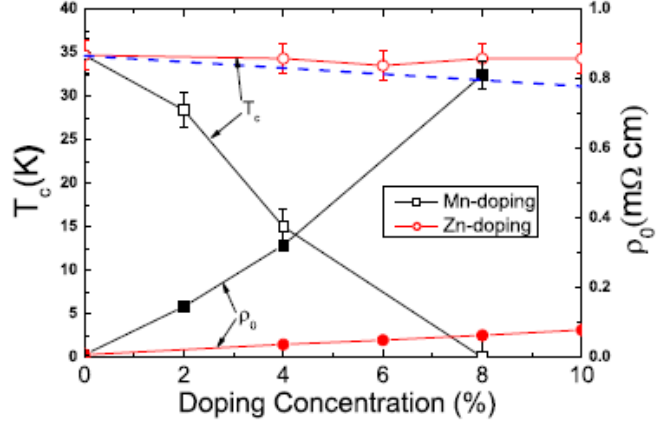

**Figure 36**. $T_c$ vs. impurity content for Ba$_{0.5}$K$_{0.5}$Fe$_{2-x}$$M_x$As$_2$ polycrystals, where $M$ = Zn and Mn, from Cheng e$t$ $al$. [51].

To elucidate the real effect of Zn, the Zn-doped $p$-type Ba$_{0.5}$K$_{0.5}$Fe$_{2-2x}$$M_{2x}$As$_2$ single crystals were synthesized by high-pressure technique as introduced in **Section 4.1**. The magnetic and nonmagnetic elements around Fe in the element periodic table were selected as the dopants, including 3$d$ metals of Mn, Co, Ni, Cu and Zn, and 4$d$ metal of Ru [49]. The decreases of $T_c$ determined by $\rho$ and $\chi$ data ($T_{c\rho}$ and $T_{c\chi}$) with the dopant concentration $x$ are given in **Figure 37**. The suppression rate by Zn was estimated at 2.22 K/% by applying a linear function fit to $T_{c\rho}$ vs. $x$, which is well accordance with the results for the BaFe$_{1.89-2x}$Zn$_{2x}$Co$_{0.11}$As$_2$ superconductors. Applying a linear function fit to $T_{c\rho}$-$x$, the suppression rates for Mn, Ru, Co, Ni, and Cu are 6.98, 0.27, 1.73, 2.21 and 2.68 K/%, respectively. Based on the density functional calculations, it was found that the impurity atoms in iron-based superconductors could be classified into three groups according to the scattering potential: (i) Mn (0.3 eV), Co (-0.3 eV), and Ni (-0.8 eV), (ii) Ru (0.1 eV) and (iii) Zn (-8 eV) [110].

Among these impurities, the nonmagnetic Zn works as a unitary scattering center due to the quite strong potential comparable to the bandwidth. Consequently, it is expected to have strong pair-breaking effect on the SC with the anisotropic SC gap. The observed robustness of SC against Zn seems to contradict the $s_{\pm}$ wave model. As discussed in **Section 2.2.2**, the reduction in $T_c$ due to the impurity is 25 K/at.%. It has also been mentioned that $T_c$ will be weakly suppressed by the impurities in the $s_{++}$ wave state in contrast to $s_{\pm}$ wave state. Among these impurities, Mn has the strongest suppression effect, even though it is much weaker than what is expected for the $s_{\pm}$ wave model (25 K/at.%). The negligible suppression effect by Ru is consistent with the results for the 1111-system [13], for which we will not discuss this isovalance-doping effect in this article.

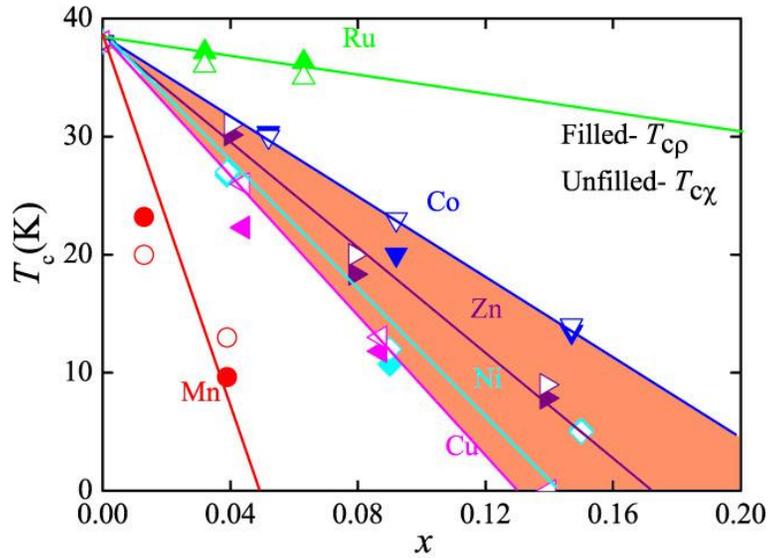

**Figure 37**. $T_c$ vs. impurity content $x$ for the $Ba_{0.5}K_{0.5}Fe_{2-2x}M_{2x}As_2$ single crystals, where $M$= Mn, Ru, Co, Ni, Cu and Zn), from Li e$t$ $al$. [49].

In addition, Tan and co-workers [58] grew Cr, Mn, Co, and Zn substituted $K_{0.8}Fe_{2-y}Se_2$ single crystals as can be seen in **Figure 38**. The $T_c$ was drastically depressed by substitution of Cr, Co, and Zn for Fe, regardless of magnetic or nonmagnetic impurities. It is interesting that all impurities showed similar suppression rate of about 30 K/at.% approaching the value expected for the $s_{\pm}$ wave model (25 K/at.%), and only about 1at.% of dopant could kill the SC completely. On the other hand, these impurities induced drastic enhancement of residual resistivity, especially for nonmagnetic Zn, which resulted in change of $\rho_0$ about 1320 $\mu\Omega$ cm just from an extremely low doping level of 0.25at.%. The dramatically large change of $\rho_0$ may indicate the occurrence of serious disorder [20, 91, 111] as analyzed in **Section 4.4**. It is rather surprising that the SC of $K_{0.8}Fe_{2-y}Se_2$ behaves as exceptionally robust against the Mn substitution which even induced a slight increase in $T_c$. The result seems likely in contrast with the Anderson theory as introduced in **Section 2.2**. The manganese atoms generally work as strongly magnetic scattering center in superconductors regardless of any order parameters, namely, it should suppress the $T_c$ for both isotropic and anisotropic pair-symmetry. In addition, substation of Mn introduces drastic charge localization and hence results in MIT in the low-$T$ region of $\rho$-$T$ curves, which was actually absent in this experiment. The authors attributed this anomalous phenomenon to the decrease of SC volume fraction owing to the nonSC islands around the Mn dopants. If so, the crystal may consist of two phases including a SC $K_{0.8}Fe_{2-y}Se_2$ and a nonSC Mn-rich phase, hence, the present results may not be considered as intrinsic substitution effect.

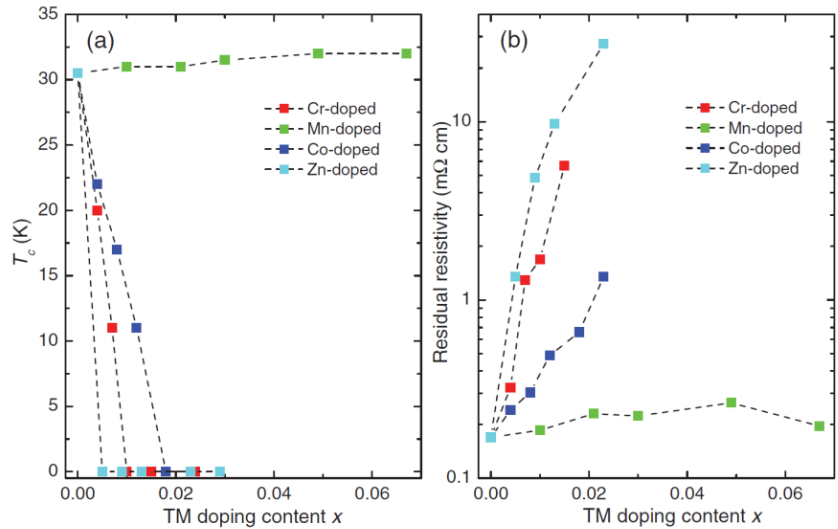

**Figure 38**. (a) The plot of the $T_c$ value vs x for the $K_{0.8}Fe_{2-y-x}M_xSe_2$ ($M$ = Cr, Mn, Co, and Zn) samples. (b) The plot of the residual resistivity vs $x$, from Tan e$t$ $al$. [58].

# 6. Pair-breaking

As discussed in **Section 2.2**, it is yet far from understanding the nature of NMI scattering based on an individual analysis of doping level dependent $T_c$. Instead, one should study various properties, among which the pair-breaking rate is the most important issue. The pair-breaking rate from NMI can be calculated as [11]

$$\alpha = \frac{z\hbar\gamma}{2\pi k_B T_{c0}} \tag{26},$$

where $\gamma$ is the electron scattering rate, $T_{c0}$ is the $T_c$ of the Zn-free compound. Based on the five-orbital model for the 122-system, theoretical study proposed a relation between $\gamma$ and $\Delta\rho_0$ as [11]

$$\Delta\rho_0 \; (\mu\Omega\text{cm}) = 0.18\gamma \; \text{(K)} \tag{27}.$$

Here $\Delta\rho_0$ is the difference of the residual resistivity between the Zn-doped and Zn-free crystals. For the $s_\pm$ wave state, SC should vanish in the range of $\alpha > \alpha^\pm_c = 0.22$ [11]. Thus, taking the results of Zn-doped BaFe$_{1.89-2x}$Zn$_{2x}$Co$_{0.11}$As$_2$ single crystals as an example, we can estimate the $\alpha$ (**Figure 39**) through

$$\alpha_1 = 0.88\frac{z\Delta\rho_0}{T_{c0}} \tag{28}$$

and using $z = 0.33$ or $0.50$,, where the effective mass for calculation of $z$ are from the ARPES experiments [54]. To obtain the elastic scattering rate, we also calculated the pair-breaking parameter using **Eq**. (17) as

$$\alpha_2 = \frac{z\hbar\gamma}{2\pi k_B T_{c0}} \tag{29}$$

by deriving $\gamma$ using the relation [11]

$$\gamma = \frac{ne^2\Delta\rho_0}{2m} \tag{30},$$

where $n$ is the carrier number estimated from the Hall data. Both $\alpha_1$ and $\alpha_2$ data change roughly linearly; thereby we applied a linear function fit to the data and estimated the critical pair-breaking parameters as 7.64, 11.49, and 6.76 for $\alpha_1$ ($z = 0.33$), $\alpha_1$ ($z = 0.05$), and $\alpha_2$, respectively. In addition, using the relation

$$\alpha_3 = \frac{\hbar \Delta \rho_0}{4\pi T_{c0}\, \mu_0\, \lambda_0^2} \qquad (31),$$

we obtained $\alpha_3 \sim 2.58$ for $\lambda = 195$ nm [86]. Obviously, the pair-breaking parameters experimentally estimated for the present system are far above the limit $\alpha^{\pm}_c = 0.22$ for the $s_\pm$ wave model, suggesting that the $s_\pm$ wave model may be not a candidate for the 122-type superconductor.

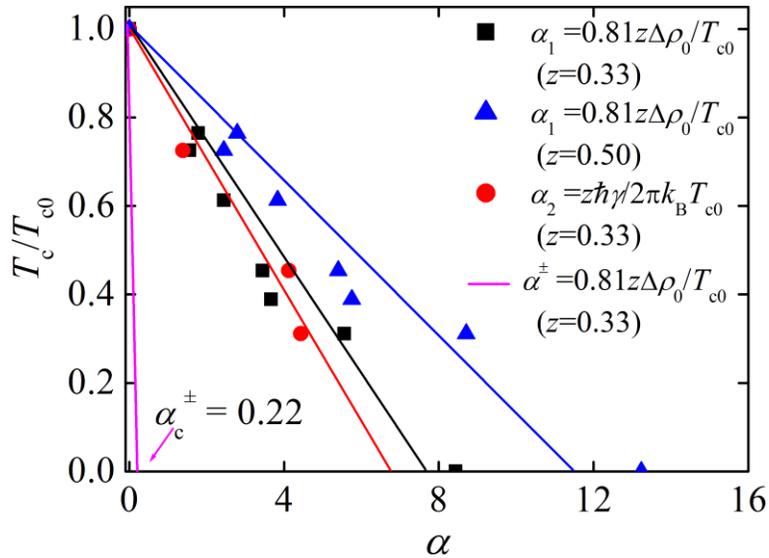

**Figure 39**. $T_c/T_{c0}$ vs. $\alpha$ with various calculations for BaFe$_{1.89-2x}$Zn$_{2x}$Co$_{0.11}$As$_2$ ($x$ = 0-0.08) by using **Eqs**. (28), (29), and (31), from Li e*t al*. [50].

It is possible that the $\alpha_1$, $\alpha_2$ and $\alpha_3$ are slightly overestimated if $\Delta\rho_0$ was overestimated due to the grain boundary or undetected electrically resistive factors. For further clarification, we made another estimation regarding the critical impurity concentration for the $s_\pm$ wave state ($n^{\pm}_{imp}$). Since Zn ($I > 1$ eV) corresponds to $n^{\pm}_{imp} \sim 0.5z/T_c$ (K), we predicted $n^{\pm}_{imp}$ to be 0.01 (0.015) for $z = 0.5$ (0.33); obviously, the experimentally determined $n_{imp}$ of 0.08 for $T_c = 0$ is much higher than the theoretical values. Thus, the quantitative discussion regarding the $n^{\pm}_{imp}$ does not accept the $s_{\pm}$ wave model for the BaFe$_{1.89}$Co$_{0.11}$As$_2$ superconductor as well.

Nakajima and co-workers [84] also calculated the pair-breaking rate on their proton-irradiated Ba(Fe$_{1-x}$Co$_x$)$_2$As$_2$ single crystals, as can be seen from **Figure 34**. Using **Eq.** (26) they found that the critical scattering rates were 6.1, 3.5, and 2.4 for the under-, optimal-, and over-doped regimes, respectively. The results are well in accordance with the chemical doping data as shown in **Figure 40**. Note that the critical pair-breaking rates of all samples are much higher than that expected for the $s_{\pm}$ pairing scenario ($\approx 0.22$).

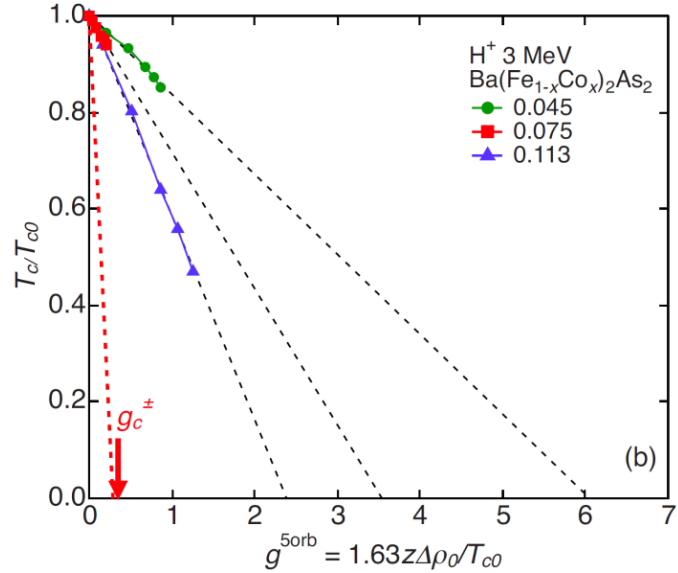

**Figure 40.** Normalized critical temperature $T_c/T_{c0}$ as a function of normalized scattering rate. The Ba(Fe,Co)$_2$As$_2$ single crystals were irradiated by proton irradiation. Dashed lines are linear extrapolations. $g_c^{\pm}$ is the critical scattering rate expected in the $s_{\pm}$ wave model, from Nakajima e*t al*. [84].

On basis of previous pair-breaking analysis for the *n*-type BaFe$_{1.89-2x}$Zn$_{2x}$Co$_{0.11}$As$_2$ superconductors, Li *et al*. also calculated the pair-breaking rate for the *p*-type Ba$_{0.5}$K$_{0.5}$Fe$_{2-2x}$$M_{2x}$As$_2$ ($M$ = Mn, Ru, Co, Ni, Cu and Zn) single crystals [49]. Here $\alpha$ can be estimated from **Eq.** (26) by using $z = 0.50$ as shown in **Figure 41**. The $T_c/T_{c0}$ changes against $\alpha$ in roughly linear way; thereby we applied a linear function fit to the data and estimated the critical pair-breaking parameters as 6.52, 5.23, 4.24, 5.41 and 6.05 for impurities of Mn, Co, Ni, Cu and Zn, respectively. The comparable result was obtained for the pair-breaking effect of Zn in the $BaFe_{1.89-2x}Zn_{2x}Co_{0.11}As_2$ system as $\alpha = 11.49$ with $z = 0.5$. The results are consistent with the proton irradiated $Ba(Fe,Co)_2As_2$ experiments [84]. Obviously, the pair-breaking parameters experimentally estimated for the present system are far above the limit of $\alpha^+_c = 0.22$ expected from the $s_\pm$ wave model, suggesting that the $s_\pm$ wave model may be unlikely in the 122-type FBS.

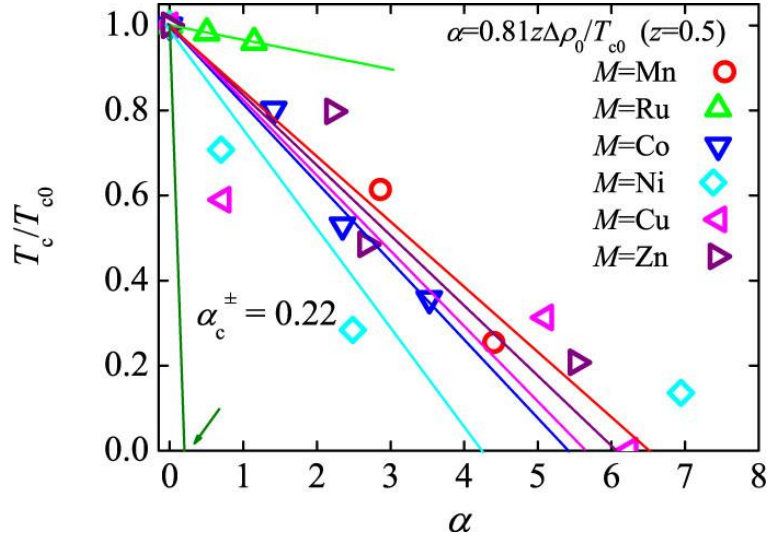

**Figure 41**. $T_c/T_{c0}$ vs. pair-breaking rate $\alpha$ for $Ba_{0.5}K_{0.5}Fe_{2-2x}M_{2x}As_2$ single crystals with $M$ = Mn, Co, Ni, Cu and Zn. The $T_c$ of each impurity-doped sample is normalized with $T_{c0}$ of the impurity-free compound. The $\alpha$ is estimated as **Eq**. (28), from Li e*t al*. [49].

To understand the relative low pair-breaking rate of Zn on the 122-type iron pnicitides, Chen and co-workers [60] calculated the disorder effects of Zn impurities in the strong (unitary) scattering limit on various properties of the system in the case of $s_\pm$ wave pairing symmetry, by solving the lattice Bogoliubov-de Gennes equation (BdG) self-consistently. **Figure 42** shows the suppression effect of Zn on the superfluid density in $BaFe_{2-2x-2y}Zn_{2x}Co_{2y}As_2$. The local superfluid density was found to decay significantly with impurity content of 3.83% at zero temperature as visualized in **Figure 42(a)**. As a result, the zero temperature bulk superfluid density $\bar{\rho}_s$ and $T_c$ are completely destroyed by ~8% of Zn impurities, implying the importance of spatial disorder-induced fluctuations. However, the decrease rate of $\bar{\rho}_s$ is larger than that of $T_c$, which were attributed to the "Swiss-cheese" scenario for a short coherence superconductor. Further corroborating with Uemura plot provides further proof for the different rates of suppression as shown in **Figure. 42(c).** The Uemura plot indicates the breakdown of the Abrikosov-Gorkov theory for impurity-averaged Green's functions. The numerical results are in agreement with the experiments, namely, SC phase is fully suppressed close to the critical impurity concentration of roughly $n_{imp} \approx 8\%$.

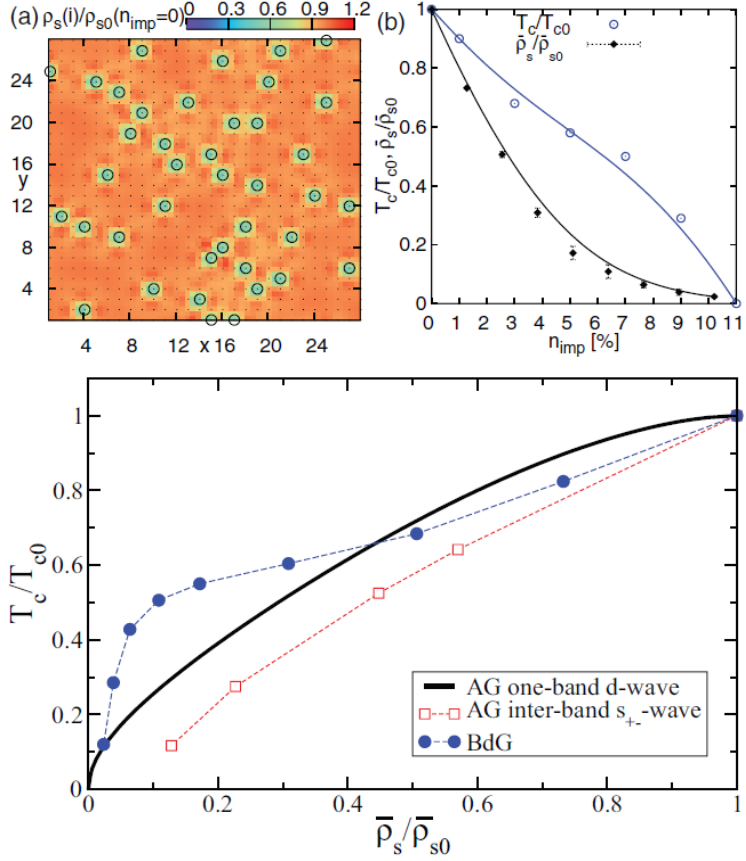

**Figure 42**. Suppression effect of Zn on the superfluid density in BaFe$_{2-2x-2y}$Zn$_{2x}$Co$_{2y}$As$_2$. (a) The intensity of normalized local superfluid density $\rho_s(\mathbf{i})/\rho_s(0)$ with Zn doping level of 3.83% at zero temperature. The open black circles indicate the impurity locations. (b) The Zn doping level dependent bulk superfluid density $\bar{\rho}_s$ at 0 K and $T_c$. The data are averaged over five randomly distributed impurity configurations. (c) The normalized $T_c/T_{c0}$ as a function of normalized $\bar{\rho}_s/\bar{\rho}_{s0}$ shown in the Uemura plots based on short-coherence superconductors. $\bar{\rho}_s$ is obtained from a pristine system. The results of the one-band Abrikosov-Gorkov calculations for $d$-wave pairing symmetry and the two-band Abrikosov-Gorkov calculations for the $s_\pm$ wave symmetry are also plotted for comparison, from Chen *et al*. [60].

For the Zn substitution results of 1111-system from Li *et al*. [112], recent simulation studies also showed the scattering rate. **Figure 43** gives the calculation of $T_c$ for Zn scattering rates and the experimental results for impurity content dependent $T_c$. They suggested that FBS might be characterized by the strength of the effective on-site pairing potential $g_0$, and proposed three types of cases for the disorder effect: (i) Large $g_0$ (**Figure 43(a)**), where the on-site pairing dominates, and $T_c$ is independent of Zn doping. The experiment results of under- and optimal-doped LaFe$_{1-y}$Zn$_y$AsO$_{1-x}$F$_x$ were believed as consistent with this case (**Figure 43(b)**). (ii) Weak $g_0$, SC is destroyed by the impurity. The authors considered the results for over-doped LaFe$_{1-y}$Zn$_y$AsO$_{1-x}$F$_x$, LaFe$_{1-x}$Zn$_x$AsO$_{1-\delta}$, and BaFe$_{2-2x-2y}$Zn$_{2x}$Co$_{2y}$As$_2$ satisfied to this model. (iii) $g_0 \approx g_2$, $T_c$ is initially suppressed rapidly with Zn-doping, and then saturates with a certain doping level. The data from polycrystalline SrFe$_{1.8-2y}$Zn$_{2x}$Co$_{0.2}$As$_2$ was classified to this case.

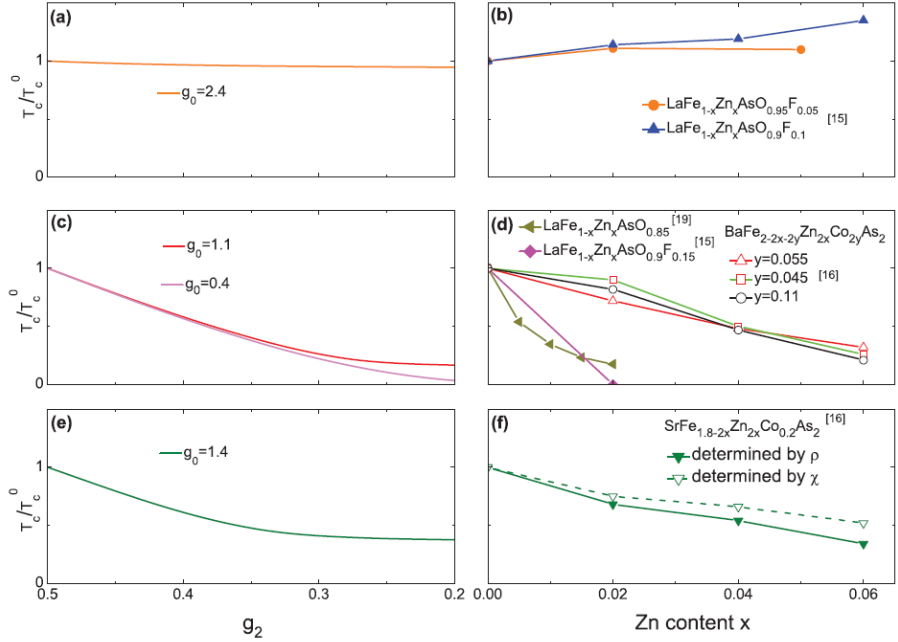

**Figure 43.** (a), (c) and (e) $T_c/T_{c0}$ vs. pair-breaking rate $g_2$ with strong, weak, and moderate on-site pairing coupling $g_0$. (a), (c) and (e): Zn-impurity effect on $T_c$ in various FBS observed in experiments. The references of the data are listed, from Yao e*t al*. [112].

For a comparison, we summarized the above data of the NMI study on single crystals in **Figure 44**. The $T_c/T_{c0}$-$\alpha$ curves locate at three different regions marked as I, II, and III in **Figure 44**(b).

Region I: For the K$_{0.8}$Fe$_{2-y-x}$Zn$_x$Se$_2$ data, although the SC was depressed with substituting only 0.25at.% of Fe-site by Zn, the change of residual resistivity $\Delta\rho_0$ was quite high as about 1320 $\mu\Omega$ cm, as a result the critical pair-breaking rate is still large as around 19.94. Nevertheless, we cannot define 19.94 as the real critical pair-breaking rate of K$_{0.8}$Fe$_{2-y-x}$Zn$_x$Se$_2$, unless one can grow more crystals with SC partially suppressed by Zn to fit linear function of $T_c/T_{c0}$ vs. $\alpha$, namely, the critical $\alpha$ may be more less than 19.94. However, one may confront with a technical difficulty during the crystal growth due to the extremely low doping level as <0.25at.%. Larger pair-breaking rates are also observed in the Zn-substituted 122-type crystals, regardless of $p$-type $Ba_{0.5}K_{0.5}Fe_{2-2x}Zn_{2x}As_2$ or $n$-type $BaFe_{1.89-2x}Zn_{2x}Co_{0.11}As_2$, whose critical $\alpha$'s are mostly as 11. Nevertheless, the evaluation of well-characterized microbridges exhibits a considerably less critical pair-breaking rate as 1.32, which should be close to the nature of the pair-breaking rate. With respect to the calculation from Efremov $et\ al.$ [14], the critical pair-breaking rate for $s_\pm$ wave model was estimated as $\Gamma_\pm^{\text{crit}}(T_{c0} = 25\,\text{K}) = 28$ with inter- to intra-band scattering ratio $\alpha \equiv u/v = 1$ from **Eq. (15)**, much larger than what was expected from Onari's model ($\alpha_c^\pm = 0.22$) [11]. In the case of Onari's model, the authors considered a "natural" formulation for impurity potential, $i.e.$, diagonal on the basis of the five Fe $d$-orbitals, automatically leading to significant interband scattering if one transforms back to the band basis. From Efremov's estimation [14], however, the ratio between inter- and intra-band scattering has been considered as the pair-breaking contribution (see the detailed discussion in **Section 2.2**), as a result the pair-breaking rates of most data collected in **Figure 44** are faster than the expected data of $\Gamma_\pm^{\text{crit}}$. The results may indicate that measurements of $T_c$ suppression with respect to the amount of chemical potential or disorder can hardly determine the gap structure in multiband systems.

Region II: The pair-breaking rates from irradiation data are quite similar for proton irradiated $BaFe_{1.85}Co_{0.15}As_2$ ($\alpha$=3.50) [84] and $\alpha$-particle irradiated $LaFeAsO_{0.7}F_{0.3}$ ($\alpha$=2.91) [82], although different irradiation sources and different superconductor systems were used. Considering the relatively larger size and power of these particles, impurities other than nonmagnetic defects may provide additional influence on the pair-breaking evaluation, as discussed in **Section 3.2**.

Region III: Under the light electron particle irradiation the 122-type crystals are induced similar critical pair-breaking rates as 1.32, 1.26 and 1.21 for $BaFeAs_{2-x}P_x$, $Ba_{1-x}K_xFe_2As_2$, and $BaFe_{1.76}Ru_{0.24}As_2$, respectively. Particularly, the pair-breaking rate of chemical doping (data from the Ba$_{1-x}$K$_x$Fe$_2$As$_2$ microbridges) seems to resemble those of electron-irradiation induced defects. Although the pair-breaking parameters of the electron-irradiated crystal and Zn-doped microbridges are a slightly larger than the expected critical value for $s_\pm$ wave ($\alpha^\pm_c = 0.22$), being about 5 times as $\alpha^\pm_c$, the experimental results are consistent with the theoretical calculation, which provide strongly evidence for $s_\pm$ wave as the nature of pair symmetry in the FBSs.

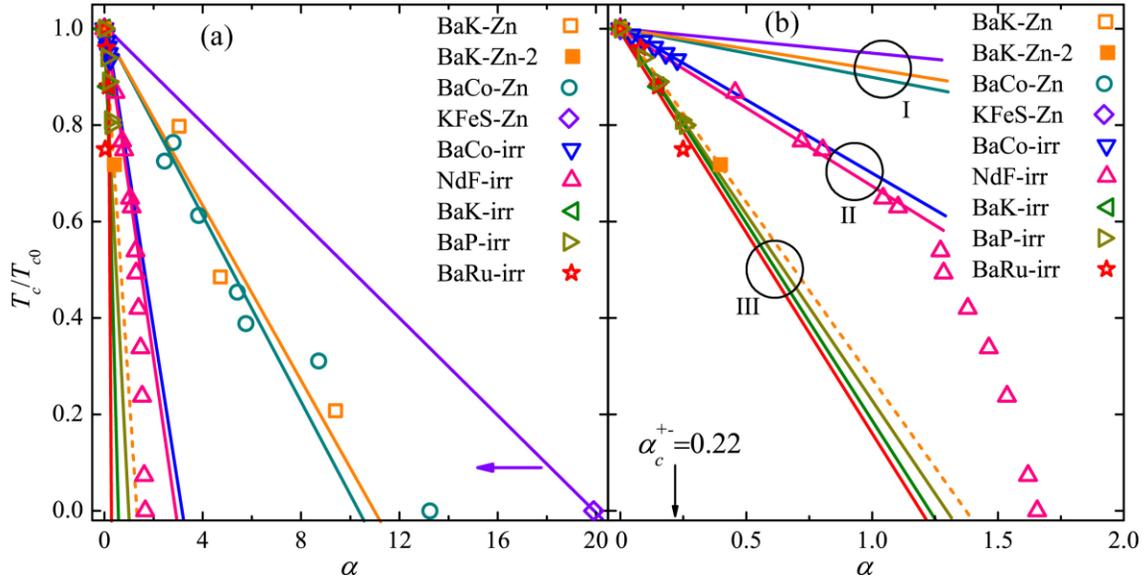

**Figure 44**. (a) $T_c/T_{c0}$ vs. $\alpha$ for various single crystals, here all samples are under the optimal-doped regimes. (b) The enlarged view on the low pair-breaking region. The pair-breaking rate $\alpha$ was estimated as **Eq**. (28) for single crystals BaK-Zn (Ba$_{0.5}$K$_{0.5}$Fe$_{2-2x}$Zn$_{2x}$As$_2$, $x$=0-0.15) [50], BaK-Zn-2 (Ba$_{0.5}$K$_{0.5}$Fe$_{2-2x}$Zn$_{2x}$As$_2$, $x$=0 and 0.05) was measured from microbridges [90], BaCo-Zn (BaFe$_{1.89-2x}$Zn$_{2x}$Co$_{0.11}$As$_2$, $x$ = 0-0.08) [49], KFeS-Zn (K$_{0.8}$Fe$_{2-y-x}$Zn$_x$Se$_2$, $x$ = 0 and 0.005) [58], BaCo-irr (BaFe$_{1.85}$Co$_{0.15}$As$_2$ under different proton-particle irradiation) [84], NdF-irr (NdFeAsO$_{0.7}$F$_{0.3}$ under different $\alpha$-particle irradiation) [82], BaK-irr (Ba$_{1-x}$K$_x$Fe$_2$As$_2$ under electron irradiation) [86], BaP-irr (BaFeAs$_{2-x}$P$_x$ under different electron irradiation) [87], and BaRu-irr (BaFe$_{1.76}$Ru$_{0.24}$As$_2$ under different electron irradiation) [88]. $\alpha^\pm_c$ is the pair-breaking parameter expected for the $s_\pm$ wave model, and it was calculated as around 0.22. All data are fitted in linear and marked in the same color as the data.

# 7. Summary

In this review, we overviewed the theoretical and experimental progress in the nonmagnetic impurity substitution study on the Fe-based superconductors. Theoretically, scientists proposed two highly potential order parameter symmetries, i. e. the multi-gap $s_{++}$ and $s_{\pm}$ wave. Because of the complex gap structure, exceptive magnetic and metallic behaviors of Fe-based superconductors, influences of nonmagnetic defects may vary in chemical potential, impurity disorder, inter- and intra-band scattering features, and electron localization, and of course the most important information for the understanding of symmetry of the superconducting order parameter. Experimentally, the first concern on the defects is naturally focused on its sample quality. Zn impurities have been successfully substituted into the single-crystalline 122-type superconductors, providing significant results by various measurements. Considering the effects of Zn impurities, the largest one is found in its immediate neighborhood within a ball region, which have not yet been investigated by using local probes such as NMR, NQR, $\mu$SR, or STM, while proofed by superconducting order-parameter phase-slip phenomenon in the Zn-doped nanowires, in which the radius of the correlation length of order around the non-superconducting regions around Zn was observed as a few lattice spacings. A single impurity also influences the electronic state and local moment, magnetic response of the $Fe_2X_2$ planes both on macroscopic scale as antiferromagnetic state and local scale of moment.

Most of experiments on the effects of nonmagnetic impurities are from the transport properties. The Zn ions provided direct modifications of the superconductivity, including the suppression of critical temperature and the enhancement of residual resistivity. Particularly, the superconductivity disappears when the residual resistance per plane accesses to the quantum resistance limit (6.45 k$\Omega$). The Hall measurements demonstrated that Zn impurities contributed weakly modification of the carrier densities. However, the Zn seems unlikely to change the slope of $T^2$ dependent Hall angle, suggesting the inducing in-plane impurity scattering rate with substitution of Zn. On the other hand, Zn ions can also influence the heat capacity, the London penetration depth, the electronic thermoelectric power, and so on.

On basis of these experiments we emphasize the qualitative feature of impurities effects on superconducting state for many FBS systems. Considering an overall nice agreement between many different experiments on different materials systems, we can safely conclude that the nonmagnetic impurity can suppress the superconductivity with major concerns about sample quality variations. As a summarization, the pair-breaking rates of nonmagnetic defects on various superconductors are collected to compare with various theoretical approaches. The pair-breaking rates of the electron-irradiated 122-type crystals and the Zn-doped microbridges are consistent with the theoretical calculated value for $s_\pm$ wave ($\alpha_c^\pm = 0.22$), which provide strongly evidence for $s_\pm$ wave as the nature of pairing symmetry in the Fe-based superconductors, but not an isotropic gap symmetry. However, we can hardly exclude a much more complex symmetry, $s+id$ wave, which behaves as strongly anisotropic gap symmetry as well. Further experimental and theoretical investigations are still required.

## Acknowledgements

We thank Drs. H.-H. Wen, H. Kontani, and Z.-X. Shi for valuable discussions. The work was supported by the National Natural Science Foundation of China (11234006, 51102188, and U1332143), Jiangsu Provincial Natural Science Fund (SBK2015040804), Fundamental Research Funds for the Central Universities, the Priority Academic Program Development of Jiangsu Higher Education Institutions, World Premier International Research Center from MEXT, the Grants-in-Aid for Scientific Research (25289233, 25289108) from JSPS, the Funding Program for World-Leading Innovative R&D on Science and Technology (FIRST Program) from JSPS, and the National Key Basic Research of China Grants (Nos. 2011CBA00111).